\documentclass[pra]{revtex4}
\usepackage{graphicx}
\usepackage{amsmath}
\usepackage{amssymb}
\usepackage{xcolor}
\begin{document}

\title{Klein tunneling through the trapezoidal potential barrier in graphene: conductance and shot noise}

\author{Gheorghe Sorin Paraoanu}

\affiliation{QTF Centre of Excellence, Department of Applied Physics, Aalto University School of Science, P.O. Box 15100, FI-00076 AALTO, Finland}

%\affiliation{}

\begin{abstract}
When a single-layer graphene sheet is contacted with metallic electrodes, tunnel barriers are formed as a result of the doping of graphene by the metal in the contact region. If the Fermi energy level is
modulated by a gate voltage, the phenomenon of Klein tunneling results in specific features in the conductance and noise. 
Here we obtain analytically exact solutions for the transmission and reflection probability amplitudes
using a trapezoidal potential barrier, allowing us to calculate the 
differential conductance and the Fano factor for a graphene sheet in the ballistic regime. We put in evidence an unexpected global symmetry - the transmission probability is the same for energies symmetric with respect to half of the barrier height. We outline a proposal for the experimental verification of these ideas using realistic sample parameters.
\end{abstract}

%\pacs{85.25.Cp, 42.65.Lm, 42.50.Dv, 03.70.+k}

\maketitle

Graphene is a flat monolayer crystal formed of carbon atoms arranged into a two-dimensional (2D) honeycomb lattice \cite{graphene1,graphene2}. This material provides a rich experimental platform for testing the predictions of relativistic quantum field theory, due to the fact that its low-energy excitations are massless Dirac fermions \cite{GeimNovoselov}. To study these quasiparticles, metallic contacts are fabricated on the graphene sheet, either as backgates or topgates. This results in the formation of tunnel barriers. Since contacts are essential for any electronics application, it is imperative to study the transport of electrons in this type of circuit.  

In non-relativistic quantum mechanics, tunneling is realized through evanescent waves and consequently the transmission probability is lower than unity, being in general dependent on the width and height of the barrier. In contradistinction, as a consequence of the peculiar energy dispersion, in graphene one expects the occurence of Klein tunneling \cite{Allain2011,Kim2011} -- where the probability of the particle exiting out on the other side of the barrier can be unity, irrespective to the height and width of the barrier. However, although in non-relativistic quantum mechanics the calculation of tunneling amplitude probabilities under various barriers is a straightforward exercise, in the relativistic case solutions have been provided only in some limiting or asymptotic cases. Relatively simple solutions have been obtained for the case of infinite-slope step potential \cite{katsnelson0,katsnelson2,katsnelson1,trauzzetel} as well as for smooth step potentials \cite{Falko2006,Huard2009}. It was also noticed that there exists a coincidence between the Fano factor obtained from Klein tunneling near the Dirac point for rectangular-barrier models \cite{beenakker} and the value of 0.3 corresponding to diffusion models in disordered metals \cite{beenakker_review}. For the case of linear-slope potentials, Sonin has obtained expressions for transmission and reflection under the assumption of an infinitely deep potential under the metallic contacts  \cite{sonin2008,sonin2009}, by building upon the solutions of the Dirac equation in an uniform electric field discovered by Sauter \cite{Sauter}. However, in all these works, the contacts are never explicitly considered, and the wavefunction of the incoming electrons is assumed to extend asymptotically the slope solution. This is obviously not the case in real samples, where the role of contact doping cannot be ignored \cite{Huard2008,kern,avouris,giovannetti}.

Experimental evidence of Klein tunneling has started first to emerge from resistance measurements of npn graphene samples at various backgate/topgate voltages \cite{huard,guinea,stander,kim}. Noise data have been acquired shortly after, and they confirmed the basic phenomenology. Shot noise measurements on single-layer graphene samples, have been first reported in diffusive samples \cite{dicarlo} (Fano factor of 0.35-0.38) and ballistic samples \cite{hakonen} (maximum Fano factor of 0.318-0.338). The latter results have confirmed the expected theoretical models based on evanescent waves \cite{beenakker} and large width-to-length ratio ($W/L \gg 1$), which predicted a minimum universal conductivity of $4e^{2}/(\pi \hbar)$ and a universal maximum Fano factor of 1/3 \cite{beenakker,katsnelson2}. The effect of decreasing $W/L$ has been further studied systematically in ballistic-sample experiments with graphene flakes on SiO$_{2}$ \cite{hakonen,danneau,danneau2}. More recently, the asymmetry of shot noise has been thoroughly investigated in samples of suspended graphene \cite{us}. Information about doping can be extracted not only from contact resistance measurements, but also from  noise measurements combined with measurements of differential conductance \cite{us}.

Here we fully solve the paradigmatic case of Klein tunneling under a trapezoidal barrier for propagating waves, providing general analytical results.  Based on these, all previous approximate results from the literature can be obtained by taking the appropriate limits. Our results can be readily used to analyze the data in any tunneling or noise experiment: in order to demonstrate this, we develop a simple and useful electrical model of doping and we show how to connect it to our parametrization. The solutions  provided here not only confirm the previous experimental results on tunneling in graphene, but also offer a unified methodology for the design of electronic circuits based on graphene. As the complexity of these circuits increases, with the use of multi-terminals, different doping substrates, and several sheets of graphene, the earlier approaches to tunneling based on asymptotic approximations will be insufficient. Instead, our approach based on analytically-exact solutions provides a straightforward, scalable approach to more sophisticated graphene-based circuits.

The paper is organized as follows. In Section I we introduce the problem, the main equations, and the key notations. In Section II we calculate the transmission and reflection probability amplitudes and we put in evidence their symmetries. In Section III we proceed to obtain the conductance and Fano factor for the incoherent and coherent cases. Finally, in Section IV we introduce the doping model and provide experimental predictions that can be tested using present-day technology. We conclude with an overview of the main results in Section V.

\section{Tunneling under the trapezoidal barrier: general considerations}

\begin{figure}
\includegraphics[width=0.59\columnwidth]{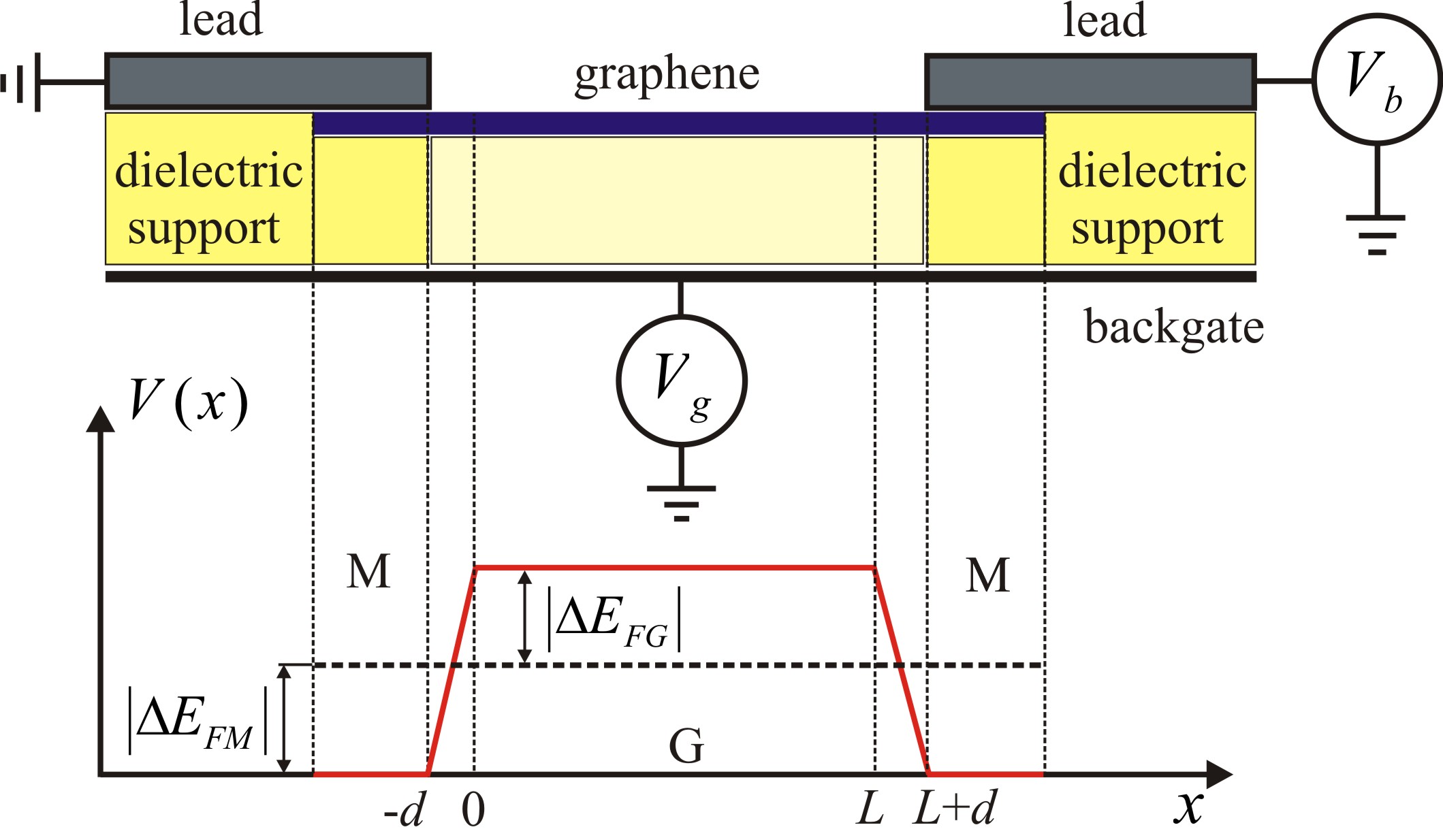}\label{general_schematic}
\caption{Schematic of a typical graphene sample and of the resulting electric potentials.}
\end{figure}

%In graphene it is possible to realize potentials that are smooth on the lattice scale $a\approx 0.2$ nm, therefore they do not induce intravalley scattering \cite{Allain2011}. 
A typical graphene sample is presented schematically in Fig. \ref{general_schematic}. The graphene is supported by a dielectric layer (or it can be suspended), and contacted by two metallic leads. A gate voltage can be applied either to the entire graphene layer (typically as a backgate) or locally (as a top gate).
The differential conductance is measured in a standard lock-in setup, with a small bias voltage applied across the structure ib order to obtain a measurable current.

Our goal is to provide a complete solution to tunneling in graphene with a trapezoidal-shape potential barrier. We present analytically exact results for the transmission and reflection coefficients, and we calculate the conductance and Fano factor for realistic experimental setups.
The model employed extends the analysis of Sonin \cite{sonin2009} to the more general case where the level of doping of graphene under the electrodes is not necessarily very large.  It assumes an electrostatic barrier of trapezoidal form, comprising five distinct regions: the middle region $0 \leq x \leq L$ with constant doping potential $V_{0}$, two slope regions $-d\leq x\leq 0$ and $L\leq x \leq L+d$, where the electric field is finite, and two regions with zero potential, $x \leq -d$ and $L+d \leq x$. Because we are interested in electrons incident on barriers, the potential is taken $y$-independent. This potential is also assumed to vary smoothly on the scale given by the graphene lattice constant of approximately $0.2$ nm, and therefore it does not induce scattering between the valleys $\mathbf{K}$ and $\mathbf{K}'$, which are separated by $|\mathbf{K} - \mathbf{K}'|$ (of the order of inverse lattice constant \cite{Allain2011}).
%We follow the notations introduced in Ref. \cite{sonin}. 
In the left slope region $-d\leq x\leq 0$ the potential is written in the form
\begin{equation}
V(x) = V_{0} + {\rm sgn \left[V_{0}\right]} a\frac{\hbar v_{\rm F}}{e} x,  \label{eq:potential}
\end{equation}
where $v_{\rm F}$ is the Fermi velocity in graphene. The sign ${\rm sgn \left[V_{0}\right]}$ realizes either a positive or negative slope. For clarity, $a$ will denote the modulus of the slope and will therefore be always a positive quantity. The potential Eq. (\ref{eq:potential}) results in a Hamiltonian that is translationally invariant in the $y$-direction, which, as detailed below,  allows us to solve the corresponding Dirac equation by the method of separation of variables.

\subsection{Dirac equation in an external potential}

Near the Dirac point  (valley) $\mathbf{K}$ the dispersion relation is linear, and as a result the electrons are described by the equation
\begin{equation}
\left[-i\hbar v_{\mathrm{F}}  \boldsymbol{\sigma}\cdot \boldsymbol{\nabla} + e V\right] \Psi_{\rm AB} = E \Psi_{\rm AB}, \label{eq:general_Dirac}
\end{equation} 
where $\boldsymbol{\sigma} = (\sigma_x , \sigma_y )$ is a vector of Pauli matrices and $v_{\rm F} \approx 10^{6}~{\rm m/s} \approx  c/300$. This is a 2D massless Dirac equation in a potential $V(x)$, where the wavefunction $\Psi_{\rm AB}(x,y) = (\psi_{\rm A}(x,y), \psi_{\rm B}(x,y))^{\mathcal{T}}$ is a pseudospin 1/2 corresponding to the two triangular sublatices A and B, and $\mathcal{T}$ stands for transpose.   The energy $E$ of the Dirac particle is measured from the same level as the potential $V(x)$. Introducing the momentum operator $\hat{\mathbf{p}} = - i \hbar \boldsymbol{\nabla}$, we can alternatively write this equation as $\hat{H}\Psi_{\rm AB} (x,y) = E \Psi_{AB} (x,y)$, where the Hamiltonian is 
$\hat{H} = \hat{\mathbf{p}}\cdot \boldsymbol{\sigma} + eV$. For clarity, in order to distinguish them from the pseudospin operators, we reserve the hat notation for operators acting on continuous variables. Since the potential $V$ is invariant in the $y$-direction, the corresponding $y$ component of the momentum is conserved, $i \hbar \dot{\hat{p}}_{y} = [\hat{p}_{y}, H] = 0$, and we can write its eigenvalue as $p_{y} = \hbar k_{y}$, where $k_{y}$ is the initial-state wavevector. Also the velocity operator is a Pauli matrix, since $\dot{\hat{\mathbf{r}}} = -(i/\hbar)[\hat{\mathbf{r}},\hat{H}]=\boldsymbol{\sigma}$, where $\mathbf{r} = (x,y)$.
The expression for the electrical current can be found from the continuity equation as \cite{Allain2011}
\begin{equation}
\mathbf{j} = e v_{\mathrm{F}}\psi^{\dag}_{\rm AB} \boldsymbol{\sigma} \psi_{\rm AB}, \label{current}
\end{equation}
and owing to the time-independence of the problem we have $\boldsymbol{\nabla} \mathbf{j} =0$. 
%On the $x$- and $y$- components the current reads
%\begin{eqnarray}
% j_{x} &=& ev_{\rm F} \Psi_{\rm AB}^{\dag} \sigma_{z} \Psi_{\rm AB} =  ev_{\rm F}[\psi_{\rm A}^{*}\psi_{\rm B} +   \psi_{\rm B}^{*}\psi_{\rm A} ], \\
% j_{y} &=& ev_{\rm F} \Psi_{\rm AB}^{\dag} \sigma_{y} \Psi_{\rm AB}  =  -iev_{\rm F}[\psi_{\rm A}^{*}\psi_{\rm B} -   \psi_{\rm B}^{*}\psi_{\rm A} ].
% \end{eqnarray}

To proceed, it is convenient to perform a change of basis in Eq. (\ref{eq:general_Dirac}), implemented by a rotation $R_{y}(-\pi /2) =\exp (i\sigma_{y}\pi/4) = \left(\mathcal{I} + i \sigma_{y}\right)/\sqrt{2}$ clockwise by $\pi/2$ around the $y$ axis. We then have $R_{y}(-\pi /2) \sigma_{x}R^{\dag}_{y}(-\pi /2) = \sigma_{z}$, and we introduce 
$\Psi (x,y) = R_{y}(-\pi /2) \Psi_{AB} (x,y)$. In this rotated frame, the Dirac equation reads
\begin{equation}
\left(-i \partial_{x}\sigma_{z} -i \partial_{y}\sigma_{y}\right) \Psi (x,y) = {\cal K} (x) \Psi (x,y) , \label{eq:Dirac_first}
\end{equation}
where we introduced a kernel ${\cal K} (x) = [E - e V(x)]/\hbar v_{\rm F}$, and $E$ is the energy of the Dirac particle. Here $\psi_{+}(x,y)$ and $\psi_{-}(x,y)$ are the two components of the spinor  $\Psi (x,y)$ defined by
$\Psi(x,y) = (\Psi_{\rm +}(x,y), \Psi_{\rm -}(x,y))^{\mathcal{T}}$. 

Furthermore, in this rotating frame the expression for the current in the $x$ direction as obtained from Eq. (\ref{current}) becomes
\begin{equation}
j_{x} = ev_{\rm F} \Psi^{\dag} \sigma_{z} \Psi  =  ev_{\rm F}(|\psi_{+}|^2 - |\psi_{-}|^2).\label{current1}
\end{equation}
%\begin{eqnarray}
%j_{x} &=& ev_{\rm F} \Psi^{\dag} \sigma_{z} \Psi  =  ev_{\rm F}[|\psi_{+}|^2 - |\psi_{-}|^2], %\label{current1}\\
%j_{y} &=& ev_{\rm F} \Psi^{\dag} \sigma_{y} \Psi =  -iev_{\rm F}[\psi_{+}^{*} \psi_{-} - \psi_{-}^{*} %\psi_{+}].\label{current2}
%\end{eqnarray}
Using this equation it immediately follows that the $x$-component $j_{x}$ of the current is constant along both $x$- and $y$- directions. The constancy in the $y$ direction $\partial_{y} j_{x} =0$ is not surprising since it follows immediately from the $y$-invariance of the potential. However, the condition 
\begin{equation}
\partial_{x} j_{x} =0
\end{equation}
 is not trivial. It implies that the x-current is the same in all the regions, a condition which we will verify explicitly throughout the paper. For convenience, we will work with solutions satisfying $|\psi_{+}|^2 - |\psi_{-}|^2 = 1$ or 
 $j_{x} = ev_{\rm F}$, corresponding to a particle of charge $e$ being transferred across the barrier from the left to right at the Fermi velocity $v_{\rm F}$.  
 %Together with $\partial_{y}j_{y}=0$ this yields $\boldsymbol{\nabla} \mathbf{j} = 0$, ensuring the conservation of probability mentioned previously.
  
From Eq. (\ref{eq:Dirac_first})  we can also put in evidence the three symmetries of the problem:

1. {\it Sublattice symmetry:} If $\Psi$ is a solution then $\sigma_{x}\Psi$ is a solution for the same problem but with a kernel $\cal{K}$ of opposite sign. This amounts to the substitutions $\psi_{\pm} \rightarrow
\psi_{\mp}$ and $\cal{K}\rightarrow - \cal{K}$.
In terms of the original lattice, this means that $\psi_{\rm A} \rightarrow -\psi_{\rm A}$ and $\psi_{\rm B} \rightarrow \psi_{\rm B}$ produces a solution corresponding to a change of sign in energy $E \rightarrow -E$, and potential $eV\rightarrow -eV$. The current $\mathbf{j}$ changes sign under this transformation.

2.  {\it Single-valley time-reversal symmetry:} A solution of the complex conjugate of Eq. 
(\ref{eq:Dirac_first}) can be obtained by $\Psi^{*}\rightarrow \sigma_{y}\Psi$ (or
$\psi_{\pm}^{*}\rightarrow \mp i \psi_{\mp}$) and the same kernel $\cal{K}$. This can be seen also from taking the complex conjugate of the Hamiltonian generating Eq. (\ref{eq:Dirac_first}), $(-i\partial_{x}\sigma_{z} -i \partial_{y} \sigma_{y})^{*} = \sigma_{y}(-i\partial_{x}\sigma_{z} -i \partial_{y} \sigma_{y})\sigma_{y}$.
In terms of the original lattice this means $\psi_{\rm A}^{*}\rightarrow -i\psi_{\rm B}$, $\psi_{\rm B}^{*}\rightarrow i \psi_{\rm A}$. The current $\mathbf{j}$ changes sign by this transformation, as expected from a time-reversed process.

3. {\it Particle-hole symmetry:} It is obtained by combining the time-reversal with the sublattice symmetry. A solution of the complex conjugate of Eq. 
(\ref{eq:Dirac_first}) can be obtained by $\Psi^{*}\rightarrow \sigma_{z}\Psi$ (or
$\psi_{\pm}^{*}\rightarrow \pm \psi_{\pm}$) and $\cal{K}\rightarrow - \cal{K}$. 
This can be seen also from taking the complex conjugate of the Hamiltonian generating Eq. (\ref{eq:Dirac_first}), $(-i\partial_{x}\sigma_{z} -i \partial_{y} \sigma_{y})^{*} = - \sigma_{z}(-i\partial_{x}\sigma_{z} -i \partial_{y} \sigma_{y})\sigma_{z}$. In terms of the original pseudospin associated with the sublattice $A$ and $B$ this transformation reads $\psi_{\rm A}^{*} \rightarrow \psi_{\rm B}$ and $\psi_{\rm B}^{*} \rightarrow \psi_{\rm A}$.  The current $\mathbf{j}$ remains invariant under this symmetry.

Since in our system the kernel ${\cal K}$ depends only on $x$, we can search for a solution of Eq. (\ref{eq:Dirac_first}) by writing $\psi_{\pm}(x,y)=\psi_{\pm}(x)\exp (ik_{y} y)$. The plane wave in the $y$-direction is appropriate for the subsequent calculation of the conductance and Fano factors, for which the standard procedure employs periodic boundary conditions. 
This results in
\begin{equation}
i \frac{d}{dx}\Psi (x) = - {\cal K}(x) \sigma_{z} \Psi (x) - i k_{y}\sigma_{x}\Psi (x).\label{eq:Dirac}
\end{equation}
Note at this point that the dimension of $a$ is [length]$^{-2}$, while that of ${\cal K}(x)$ and of $k_{y}$ is [length]$^{-1}$. We have thus obtained a fully adimensional Dirac equation Eq. (\ref{eq:Dirac}), with all the physical constants embedded in the kernel $\mathcal{K}$.

Next, we introduce a few other useful notations for the height of the Fermi level measured from the Dirac point for the graphene under the metal, which we denote as
$\Delta E_{\rm FM}$, and the same quantity for the free graphene, which we denote by $\Delta E_{\rm FG}$, see Fig. \ref{general_schematic}. We then define adimensional quantities \cite{sonin2008,us}, namely the reduced coordinate $\xi = x\sqrt{a}$,
the reduced wave-vector component in the $y$-direction $\kappa =k_{y}/\sqrt{a}$, and the reduced Fermi momentum for the free graphene,
\begin{equation}
v=\frac{\Delta E_{\rm FG}}{\sqrt{a}\hbar v_{\rm F}}.
\end{equation}
Also we introduce the height of the potential barrier by
\begin{equation}
eV_{0} = \Delta E_{\rm FM}- \Delta E_{\rm FG},
\end{equation}
and the finite-slope impact parameter
\begin{equation}
p_{0} = \frac{eV_{0}}{\hbar v_{\rm F}\sqrt{a}}.
\end{equation}
%With these notations, the momentum of the Dirac particles in the contact region is
%\begin{equation}
%\frac{\Delta E_{\rm FM}}{\hbar v_{\rm F}}=\sqrt{a}(p_{0} + v).
%\end{equation}
It is also natural to define a reduced kernel $\mathbb{K} (\xi ) = \mathcal{K}(x)/\sqrt{a}$. 
We can now express Eq. (\ref{eq:Dirac}) in the variable $\xi$ on the $\pm$ components
\begin{eqnarray}
i\frac{d}{d\xi }\psi_{+}(\xi ) + i \kappa \psi_{-}(\xi ) &=& -\mathbb{K} (\xi )\psi_{+}(\xi), \label{11}\\
i\frac{d}{d\xi }\psi_{-}(\xi ) + i \kappa \psi_{+}(\xi ) &=& \mathbb{K}(\xi)\psi_{-}(\xi), \label{12}
\end{eqnarray}
obtaining two equations fully written in terms of the reduced parameters.

\section{Transmission coefficients}

With the notations and setup above we are now in a position to solve analytically the problem of tunneling though propagating modes of the Dirac particle across the trapezoidal barrier.

\subsection{General expressions}

\subsubsection{\bf Top of the barrier, $\mathbf{x>0}$}

In this region the reduced kernel reads
%\begin{equation}
%{\cal K}(x|x>0) = \frac{1}{\hbar v_{\rm F}}\Delta E_{\rm FG} = \sqrt{a}v,
%\end{equation}
\begin{equation}
\mathbb{K}(\xi )_{\xi >0} = v,
\end{equation}
for both positive and negative $\Delta E_{\rm FG}$.
%\begin{equation}
%\Psi (\xi)_{x>0}= \left( \begin{array}{c} \frac{1}{2} + \frac{\sqrt{v^2 - \kappa^2} +i {\rm sgn}[v] %\kappa}{2\left| v \right|} \\ -\frac{1}{2} + \frac{\sqrt{v^2 - \kappa^2} + i{\rm sgn}[v]\kappa}{2\left| v %\right|} \end{array} \right) \sqrt{\frac{\left| v \right|}{\sqrt{v^2 - \kappa^2}}}e^{i {\rm sgn}[v] %\sqrt{v^2 - \kappa^2}\xi},
%\end{equation}
The solution for a particle moving in the direction of the positive $x$ axis is
\begin{equation}
\Psi (\xi)_{x>0}= \left[ \begin{array}{c} \psi_{+}(v,\kappa) \\ \psi_{-}(v,\kappa)\end{array} \right]e^{i {\rm sgn}[v] \sqrt{v^2 - \kappa^2}\xi},
\end{equation}
where
\begin{equation}
\psi_{\pm}(v,\kappa) = \frac{\pm |v| + \sqrt{v^2 - \kappa^2} + i {\rm sgn}[v]\kappa}{2\sqrt{|v|\sqrt{v^2 - \kappa^2}}}.
\end{equation}
It is straightforward to calculate the $j_{x}$ current, and we find from the equations above and employing Eq. (\ref{current1}) that $j_{x} = e v_{\rm F}$.
%,\ref{current2}) 
%\begin{eqnarray}
%j_{x} &=& e v_{\rm F}, \\
%j_{y} &=& ev_{\rm F}\frac{{\rm sgn}[v] \kappa}{\sqrt{v^2 - \kappa^2}},
%\end{eqnarray}
We note that in this region the momentum in the $x$-direction is $p_{x, \rm FG} = {\rm sgn} [\Delta E_{\rm FG}]\sqrt{(\Delta E_{\rm FG}/v_{\rm F})^2
- \hbar^2 k_{y}^2}$ and it is a real quantity. This implies  $|\Delta E_{\rm FG}| \geq \hbar v_{\mathrm{F}}k_{y}$, or  $|v|\geq\kappa$.

\subsubsection{\bf Region under the metal $\mathbf{x<-d}$}

For the region of graphene under the  metal we have
%\begin{equation}
%{\cal K}(x|x<-d)= \frac{1}{\hbar v_{\rm F}}\Delta E_{\rm FM} = \sqrt{a}(p_{0}+v),
%\end{equation}
\begin{equation}
\mathbb{K}(\xi )_{x<-d}=p_{0}+v.
\end{equation}
The solution can be expressed as a superposition between a transmitted and a reflected wave
\begin{equation}
\Psi (x)_{x<-d} = \frac{1}{t}\Psi^{(t)} (x) + \frac{r}{t}\Psi^{(r)} (x), \label{eq:overall}
\end{equation}
with $r$ and $t$ denoting respectively the reflection and transmission coefficients.

In terms of the variable $\xi$ we write these waves as
%\begin{equation}
%\Psi^{(t)}(\xi) =\left(\begin{array}{c} \frac{1}{2} + \frac{\sqrt{(p_{0}+v)^2 - \kappa^2} +i{\rm sgn}
%[p_{0}+v] \kappa}{2\left| p_{0} +v \right|} \\ -\frac{1}{2} +
%\frac{\sqrt{(p_{0}+v)^2 - \kappa^2}+
%i{\rm sgn}
%[p_{0}+v]\kappa}{2\left| p_{0} +v \right|} \end{array}\right) \sqrt{\frac{\left| p_{0} +v %\right|}{\sqrt{(p_{0}+v)^2 - \kappa^2}}}e^{+ i {\rm sgn}
%[p_{0}+v]\sqrt{(p_{0}+v)^2 - \kappa^2}\xi}, 
%\end{equation}
\begin{equation}
\Psi^{(t)}(\xi) = \left[\begin{array}{c} \psi_{+}(p_{0}+v,\kappa) \\ \psi_{-}(p_{0}+v,\kappa) \end{array}\right]
e^{+ i {\rm sgn}
	[p_{0}+v]\sqrt{(p_{0}+v)^2 - \kappa^2}\xi},
\end{equation}
and
\begin{equation}
\Psi^{(r)}(\xi) = -\left[\begin{array}{c} \psi_{-}^{*}(p_{0}+v,\kappa) \\ \psi_{+}^{*}(p_{0}+v,\kappa) \end{array}\right]
e^{- i {\rm sgn}
	[p_{0}+v]\sqrt{(p_{0}+v)^2 - \kappa^2}\xi} .\label{admixt}
\end{equation}

%\begin{equation}
%\Psi^{(r)} (\xi) = \left(\begin{array}{c} \frac{1}{2} + \frac{-\sqrt{(p_{0}+v)^2 - \kappa^2} +i{\rm sgn}
%[p_{0}+v]\kappa}{2 \left| p_{0} +v \right|} \\ -\frac{1}{2} + \frac{-\sqrt{(p_{0}+v)^2 - \kappa^2} + i{\rm sgn}
%[p_{0}+v]\kappa}{2 \left| p_{0} +v \right|}\end{array}\right) \sqrt{\frac{\left| p_{0} +v \right|
%}{\sqrt{(p_{0}+v)^2 - \kappa^2}}}e^{- i {\rm sgn}
%[p_{0}+v]\sqrt{(p_{0}+v)^2 - \kappa^2}\xi}. \label{admixt}
%\end{equation}

Note also that in Eq. (\ref{admixt}) the reflected component is normalized to a negative current in the x-direction equal to $-ev_{\rm F}$ while the transmitted component is normalized to a positive $+ev_{\rm F}$.
Specifically,
\begin{equation}
j^{(t)}_{x}=-j^{(r)}_{x}= ev_{\rm F}.
\end{equation}
Moreover, one can explicitly check that the current corresponding to the total wavefunction Eq. (\ref{eq:overall})
is $+ev_{\rm F}$, that is 
\begin{equation}
j_{x} =  ev_{\rm F},
\end{equation}
given that that $|r|^2 + |t|^2=1$. Thus the correct normalization is recovered in this region. 
%It also means that there is no interference in the $x$-direction between the transmitted and reflected waves. For the $y$-current we find
%\begin{equation}
%j^{(t)}_{y} = j^{(r)}_{y} = ev_{\rm F} \frac{{\rm sgn} [p_{0} +v]\kappa}{ (p_{0}+v)^2 -\kappa^2},
%\end{equation}
%however in this case the total current present interference terms.
Similarly to the previous region, the condition  $|\Delta E_{\rm FM}| \geq \hbar v_{\mathrm{F}}k_{y}$ must hold, ensuring that the momentum in the $x$-direction $p_{x, \rm FM} = {\rm sgn}[\Delta E_{\rm FM}]\sqrt{(\Delta E_{\rm FM}/v_{\rm F})^2 - \hbar^2 k_{y}^2}$ is real.
This results in $|p_{0}+v|>\kappa$.

\subsubsection{\bf  Slope of the barrier, $\mathbf{-d<x<0}$}

In this region we have
%\begin{equation}
%{\cal K}(x|-d<x<0) = {\rm sgn}[V_{0}] a(x-x_{0}),
%\end{equation}
\begin{equation}
\mathbb{K}(\xi)_{-d<x<0} = - {\rm sgn}[V_{0}] (\xi - \xi_{0}),
\end{equation}
where  $\xi_{0} = \sqrt{a} x_{0}$ is the crossing point between the slope and the Fermi level. The solutions in this case have been found by Sauter \cite{Sauter} and
are linear combinations of
\begin{equation}
\left[\begin{array}{c} F(\xi - \xi_{0}, \kappa) \\ G (\xi - \xi_{0} , \kappa)
 \end{array} \right] {\rm and}
\left[\begin{array}{c} G^{*}(\xi - \xi_{0}, \kappa ) \\ F^{*}(\xi - \xi_{0}, \kappa )
\end{array} \right] ,
\end{equation}
in the case of positive slope, and the conjugate of these expressions for negative slope.
%The fact that the negative-slope solution is the conjugate of the positive-slope solution follows from general time-reversal considerations: if the motion of a particle on a positive slope, when time-reversed, will look like motion on a negative-slope. In this case, from the time-dependent Dirac equations we can see that the solutions are the complex conjuate of each other.
Here $F$ and $G$ are given by  
\begin{equation}
F(\xi- \xi_{0}, \kappa) = e^{-i(\xi- \xi_{0})^2 /2}M\left(-\frac{i\kappa^2}{4}, \frac{1}{2}, i(\xi- \xi_{0})^2\right),
\end{equation}
and
\begin{equation}
G(\xi-\xi_{0}, \kappa) = - \kappa (\xi- \xi_{0}) e^{-i(\xi-\xi_{0})^2/2}M\left(1- \frac{i\kappa^2}{4}, \frac{3}{2}, i(\xi- \xi_{0})^2 \right).
\end{equation}
where $M$ is the Kummer confluent hypergeometric function.
In Appendix \ref{kummer} we give a derivation of these expressions.
Importantly, $|F(\xi, \kappa)|^2 - |G(\xi, \kappa)|^2 =1$,  which, as before, turns  out to be essential for obtaining the correct x-current.
In spite of the fact that the negative-slope solution is just the conjugate of the positive-slope solution, the intersection points $x_{0}$, which contain the information about the position of the Fermi level, have opposite sign, and to avoid any confusion the two cases have to be considered separately.

\paragraph{\bf Positive slope at the left side of the barrier}

\begin{figure}
\includegraphics[width=0.49\columnwidth]{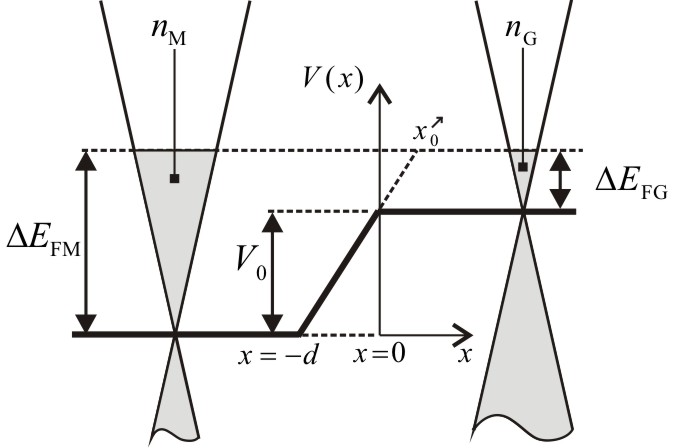}
\caption{Schematic of the potential at the left side of the barrier with positive slope.}\label{left_positive_barrier}
\end{figure}

We now look at the situation when the slope of the potential is positive at the left side of the barrier.
 As a convention, in the following $a$ will be defined as a positive number, $a>0$. The height of the electrical potential is positive, $V_{0}\equiv \Delta E_{\rm FM}-\Delta E_{\rm FG}>0$. A schematic of the potential experienced by the Dirac fermions is shown in Fig. \ref{left_positive_barrier}.
The type of junction formed depends on the carrier type
in the ``metal'' $\rm M$ and ``graphene'' $\rm G$ region. We get a $n-n$ junction if $\Delta E_{\rm FM}>0$ and $\Delta E_{\rm FG}>0$, a $n-p$ junction if $\Delta E_{\rm FM}>0$ and $\Delta E_{\rm FG}<0$, and finally a $p-p$ junction if $\Delta E_{\rm FM}<0$ and $\Delta E_{\rm FG}<0$. Note that $p-n$ junctions cannot form for positive slopes. 
The potential has the form
\begin{equation}
V(x) = V_{0}+a \frac{\hbar v_{F}}{e}x,
\end{equation}
resulting in the useful relation
\begin{equation}
\sqrt{a} d = p_{0},
\end{equation}
when $V (-d)=0$ is used.
The kernel on the slope of the barrier is given by
\begin{equation}
{\cal K}(x) = \frac{\Delta E_{\rm FM} - e V(x)}{\hbar v_{\rm F}} = \frac{\Delta E_{\rm FG}}{\hbar v_{\rm F}} -a x = -a (x-x_{0}^{(+)}),
\end{equation}
where the crossing point is $x_{0}^{\nearrow}= \Delta E_{\rm FG}/a \hbar v_{\rm F}=v/\sqrt{a}$, or $\xi^{\nearrow}_{0} = \sqrt{a}x_{0}^{\nearrow} = v$, where the superscript $\nearrow$ denotes a positive slope.
Thus the crossing point takes positive values for $\Delta E_{\rm FG}>0$ and negative values for  $\Delta E_{\rm FG}<0$.

The general solutions are
\begin{equation}
\left[\begin{array}{c} \psi_{+}^{\nearrow}(\xi ) \\ \psi_{-}^{\nearrow}(\xi ) \end{array} \right]_{-d <x <0}
=C_{1}^{\nearrow}(v,\kappa ) \left[\begin{array}{c} F(\xi - v, \kappa) \\ G(\xi - v ,\kappa )
 \end{array} \right]
 +
 C_{2}^{\nearrow} (v,\kappa )\left[\begin{array}{c} G^{*}(\xi - v, \kappa) \\ F^{*}(\xi - v, \kappa )
 \end{array} \right].
 \end{equation}
From the boundary conditions at $x=0$ we obtain
%\begin{widetext}
%\begin{eqnarray}
%C_{1}^{\nearrow} &=& \frac{\left| v \right| + \sqrt{v^2 - \kappa^2} + i {\rm sgn}[v] \kappa}
%{2\sqrt{\left| v \right|\sqrt{v^2 - \kappa^2}}}F^{*}%(-v, \kappa)+
%\frac{\left| v \right| - \sqrt{v^2 - \kappa^2} - i {\rm sgn}[v] \kappa
%}{2\sqrt{\left| v \right|\sqrt{v^2 - \kappa^2}}}G^{*}(-v, \kappa ) , \nonumber \\
%C_{2}^{\nearrow} &=& \frac{-\left| v \right| + \sqrt{v^2 - \kappa^2} + i {\rm sgn}[v] \kappa}
%{2\sqrt{\left| v \right|\sqrt{v^2 - \kappa^2}}}F(-v, \kappa)-
%\frac{\left| v \right| + \sqrt{v^2 - \kappa^2} + i {\rm sgn}[v] \kappa
%}{2\sqrt{\left| v \right|\sqrt{v^2 - \kappa^2}}}G(-%v, \kappa ) . \nonumber
%\end{eqnarray}
%\end{widetext}
%\begin{widetext}
\begin{eqnarray}
C_{1}^{\nearrow}(v,\kappa ) &=& \psi_{+}(v,\kappa )F^{*}(v, \kappa) + \psi_{-}(v,\kappa )G^{*}(v, \kappa), \\
	C_{2}^{\nearrow}(v,\kappa ) &=& \psi_{-}(v,\kappa )F (v, \kappa) + \psi_{+}(v,\kappa )G(v, \kappa).\nonumber
	\end{eqnarray}
%\end{widetext}
Importantly, the consequence of $j_{x}=ev_{\rm F}$ or $|\psi_{+}^{\nearrow}(\xi )|^2 - |\psi_{-}^{\nearrow}(\xi )|^2=1$ is the fact that $|C_{1}^{\nearrow}
(v,\kappa )|^2
-|C_{2}^{\nearrow}(v, \kappa )|^2 = 1$.

From the boundary conditions at $x=-d$ we get
%\begin{widetext}
%\begin{eqnarray}
%& &\frac{1}{t^{\nearrow}}e^{-i p_{0} {\rm sgn}[p_{0}+v ]\sqrt{(p_{0}+ v)^2 - \kappa^2}} = \\ 
%& &\frac{\left| p_{0} +v \right| + \sqrt{(p_{0}+ v)^2 - \kappa^2} - i {\rm sgn}[p_{0}+v ] \kappa
%}{2\sqrt{\left| p_{0} +v \right|\sqrt{(p_{0}+ v)^2 - %\kappa^2}}}[C_{1}^{\nearrow} F(-p_{0}-v, \kappa) + C_{2}^{\nearrow} G^{*}(-p_{0}-v, \kappa )] \\ 
%& & +
%\frac{\left| p_{0} +v \right| - \sqrt{(p_{0}+ v)^2 - \kappa^2} + i {\rm sgn}[p_{0}+v] \kappa
%}{2\sqrt{\left| p_{0} +v \right|\sqrt{(p_{0}+ v)^2 - \kappa^2}}}[C_{1}^{\nearrow} G(-p_{0}-v, \kappa ) + C_{2}^{\nearrow} F^{*}(-p_{0}-v, \kappa )],
%\end{eqnarray}
%\end{widetext}
%\begin{widetext}
\begin{eqnarray}
	& &\frac{1}{t^{\nearrow}}e^{-i p_{0} {\rm sgn}[p_{0}+v ]\sqrt{(p_{0}+ v)^2 - \kappa^2}} = \nonumber\\ 
	& & C_{1}^{\nearrow}(v,\kappa )[\psi_{+}^{*}(p_{0}+v,\kappa) F(p_{0}+v, \kappa) + \psi_{-}^{*}(p_{0}+v,\kappa)G(p_{0}+v,\kappa )] \nonumber \\
& &	-C_{2}^{\nearrow}(v, \kappa )[\psi_{+}^{*}(p_{0}+v,\kappa)G^{*}(p_{0}+v, \kappa ) + \psi_{-}^{*}(p_{0}+v,\kappa)F^{*}(p_{0}+v, \kappa )], \label{tup}
\end{eqnarray}
%\end{widetext}
and for the reflection coefficient we obtain
%\begin{widetext}
%\begin{eqnarray}
%& & \frac{r^{\nearrow}}{t^{\nearrow }}e^{-i {\rm sgn}[p_{0}+v ] p_{0} \sqrt{(p_{0}+ v)^2 - \kappa^2}} = \\
%& & \frac{-\left| p_{0} +v \right| + \sqrt{(p_{0}+ v)^2 - \kappa^2} + i {\rm sgn}[p_{0}+v] \kappa
%}{2\sqrt{\left| p_{0} +v \right|\sqrt{(p_{0}+ v)^2 - \kappa^2}}}[C_{1}^{\nearrow} F(-p_{0}-v, \kappa) + C_{2}^{\nearrow} G^{*}(-p_{0}-v, \kappa )]   \nonumber \\
%& & -
%\frac{\left| p_{0} +v \right| + \sqrt{(p_{0}+ v)^2 - \kappa^2} + i {\rm sgn}[p_{0}+v] \kappa
%}{2\sqrt{\left| p_{0} +v \right|\sqrt{(p_{0}+ v)^2 - \kappa^2}}}[C_{1}^{\nearrow} G(-p_{0}-v, \kappa) + C_{2}^{\nearrow} F^{*}(-p_{0}-v, \kappa )] . \nonumber
%\end{eqnarray}
%\end{widetext}
%\begin{widetext}
	\begin{eqnarray}
	& & \frac{r^{\nearrow}}{t^{\nearrow }}e^{i {\rm sgn}[p_{0}+v ] p_{0} \sqrt{(p_{0}+ v)^2 - \kappa^2}} = \nonumber\\
	& & C_{1}^{\nearrow}(v,\kappa )[\psi_{-}(p_{0}+v,\kappa)F(p_{0}+v, \kappa) + \psi_{+}(p_{0}+v,\kappa)G(p_{0}+v, \kappa)] \nonumber \\
& & 	- C_{2}^{\nearrow}(v, \kappa )[\psi_{+}(p_{0}+v,\kappa)F^{*}(p_{0}+v, \kappa ) + \psi_{-}(p_{0}+v)G^{*}(p_{0}+v, \kappa )]. \label{rup}
	\end{eqnarray}
%\end{widetext}
This transmission coefficient is presented in Fig.\ref{fig:transmisssions} a).

\paragraph{\bf Negative slope at the left side of the barrier}

The case of a potential with negative slope has to be treated separated. The reason for this is that the slope appears under square root in the substitutions needed to get the Sauter solution. With our convention, $a$ is  defined as a positive number, $a>0$, and the height of the electrical potential is become negative, $V_{0}\equiv \Delta E_{\rm FM}-\Delta E_{\rm FG}<0$. A schematic of the potential experienced by the Dirac fermions is shown 
in Fig. 3.
\begin{figure}
\includegraphics[width=0.49\columnwidth]{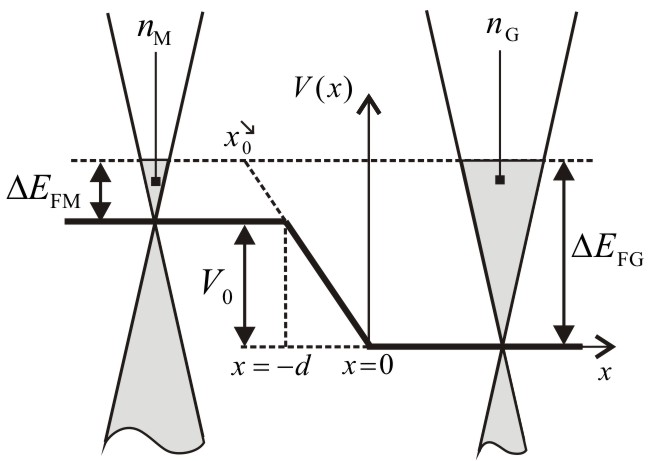}\label{left_negative_barrier}
\caption{Schematic of the potential at the left side of the barrier with negative slope.}
\end{figure}
The type of junction formed depends on the carrier type
in the ``metal'' $\rm M$ and ``graphene'' $\rm G$ region, which is set by the value of the Fermi level.  We get a $n-n$ junction if $\Delta E_{\rm FM}>0$ and $\Delta E_{\rm FG}>0$, a $p-n$ junction if $\Delta E_{\rm FM}<0$ and $\Delta E_{\rm FG}>0$, and finally a $p-p$ junction if $\Delta E_{\rm FM}<0$ and $\Delta E_{\rm FG}<0$. For negative slopes, $n-p$ junctions cannot form.
The potential has the expression
\begin{equation}
V(x) = V_{0}-a \frac{\hbar v_{F}}{e}x,
\end{equation}
resulting in 
\begin{equation}
\sqrt{a}d=-p_{0}.
\end{equation}
For the kernel on the slope of the barrier we have
\begin{equation}
{\cal K}(x) = \frac{\Delta E_{\rm FM} - e V(x)}{\hbar v_{\rm F}} = \frac{\Delta E_{\rm FG}}{\hbar v_{\rm F}} + a x = a (x-x_{0}^{\searrow}),
\end{equation}
where the crossing point is $x_{0}^{\searrow}= -\Delta E_{\rm FG}/a \hbar v_{\rm F}$, or
$\xi_{0}^{\searrow} = -\sqrt{a}x_{0}^{\searrow}=-v$,  with the superscript $\searrow$ denoting a negative slope.
 Opposite to the situation of positive slope, here the crossing point has negative values for positive $\Delta E_{\rm FG}>0$ and negative values for  $\Delta E_{\rm FG}<0$.

The general solutions are
\begin{equation}
\left[\begin{array}{c} \psi_{+}^{\searrow}(\xi ) \\ \psi_{-}^{\searrow}(\xi ) \end{array} \right]_{-d<x<0}
=C_{1}^{\searrow}(v, \kappa ) \left[\begin{array}{c} F^{*}(\xi + v, \kappa) \\ G^{*}(\xi + v , \kappa )
 \end{array} \right]
 +
 C_{2}^{\searrow }(v, \kappa ) \left[\begin{array}{c} G(\xi + v, \kappa ) \\ F(\xi + v, \kappa )
 \end{array} \right],
 \end{equation}
Again, the consequence of $j_{x}=ev_{\rm F}$ or $|\psi_{+}^{\searrow}(\xi )|^2 - |\psi_{-}^{\searrow}(\xi )|^2=1$ is the fact that $|C_{1}^{\searrow}(v,\kappa )|^2
-|C_{2}^{\searrow}(v, \kappa )|^2 = 1$.

From the boundary conditions at $x=0$ we obtain
%\begin{widetext}
%\begin{eqnarray}
%C_{1}^{\searrow}(v,\kappa ) &=& \frac{\left| v \right| + \sqrt{v^2 - \kappa^2} + i {\rm sgn}[v] \kappa
%}{2\sqrt{\left| v \right|\sqrt{v^2 - \kappa^2}}}F(v, \kappa)+
%\frac{\left| v \right| - \sqrt{v^2 - \kappa^2} - i {\rm sgn}[v] \kappa
%}{2\sqrt{\left| v \right|\sqrt{v^2 - \kappa^2}}}G(v, \kappa ) , \\
%C_{2}^{\searrow} (v, \kappa )&=& \frac{-\left| v \right| + \sqrt{v^2 - \kappa^2} + i {\rm sgn}[v]
%\kappa}{2\sqrt{
%\left| v \right|\sqrt{v^2 - \kappa^2}}}F^{*}(v, \kappa )-
%\frac{\left| v \right| + \sqrt{v^2 - \kappa^2} + i {\rm sgn}[v] \kappa
%}{2\sqrt{\left| v \right|\sqrt{v^2 - \kappa^2}}}G^{*}(v, \kappa ) ,
%\end{eqnarray}
%\end{widetext}
%\begin{widetext}
	\begin{eqnarray}
	C_{1}^{\searrow}(v,\kappa ) &=& \psi_{+}(v,\kappa )
	F(v, \kappa) - \psi_{-}(v,\kappa)G(v, \kappa ) , \\
	C_{2}^{\searrow} (v, \kappa )&=& \psi_{-}(v,\kappa )F^{*}(v, \kappa )
	-\psi_{+}(v,\kappa )G^{*}(v, \kappa ) ,
	\end{eqnarray}
%\end{widetext}
and from the boundary conditions at $x=-d$ we get
%\begin{widetext}
%begin{eqnarray}
%& & \frac{1}{t^{\searrow}}e^{i {\rm sgn}[p_{0}+v ]p_{0}\sqrt{(p_{0}+ v)^2 - \kappa^2}} = \\
%& & \frac{\left| p_{0} +v \right| +\sqrt{(p_{0}+ v)^2 - \kappa^2} - i {\rm sgn}[p_{0}+ v] \kappa
%}{2\sqrt{\left| p_{0} +v \right|\sqrt{(p_{0}+ v)^2 - \kappa^2}}}[C_{1}^{\searrow} F^{*}(p_{0}+v , \kappa ) + C_{2}^{\searrow} G(p_{0}+v, \kappa)] \nonumber \\ 
%&  & 
%+ \frac{\left| p_{0} + v \right| - \sqrt{(p_{0}+ v)^2 %- \kappa^2} + i {\rm sgn}[p_{0}+v] \kappa
%}{2\sqrt{\left| p_{0} +v \right|\sqrt{(p_{0}+ v)^2 - \kappa^2}}}[C_{1}^{\searrow} G^{*}(p_{0}+v, \kappa ) + C_{2}^{\searrow} F(p_{0}+v, \kappa)] \nonumber
%\end{eqnarray}\end{widetext}
%\begin{widetext}
	\begin{eqnarray}
& & \frac{1}{t^{\searrow}}e^{i {\rm sgn}[p_{0}+v ]p_{0}\sqrt{(p_{0}+ v)^2 - \kappa^2}} =\nonumber\\
& &  C_{1}^{\searrow}(v,\kappa )[\psi_{+}^{*}(p_{0}+v,\kappa)F^{*}(p_{0}+v , \kappa ) -
	\psi_{-}^{*} (p_{0} + v,\kappa ) G^{*}(p_{0}+v, \kappa )] \nonumber \\
& & + C_{2}^{\searrow}(v,\kappa )[\psi_{+}^{*}(p_{0}+v,\kappa )G(p_{0}+v, \kappa) - \psi_{-}^{*}(p_{0}+v,\kappa) F(p_{0}+v, \kappa)], \label{tdown}
\end{eqnarray}%\end{widetext}
and the reflection coefficient
%\begin{widetext}
%\begin{eqnarray}
%& & \frac{r^{\searrow}}{t^{\searrow}}e^{i{\rm sgn}[p_{0}+v ] p_{0} \sqrt{(p_{0}+ v)^2 - \kappa^2}} = \\
%& & \frac{-\left| p_{0} +v \right| + \sqrt{(p_{0}+ v)^2 - \kappa^2} + i {\rm sgn}[p_{0}+v] \kappa
%}{2\sqrt{\left| p_{0} +v \right|\sqrt{(p_{0}+ v)^2 - \kappa^2}}}[C_{1}^{\searrow} F^{*}(p_{0}+v, \kappa ) + C_{2}^{\searrow} G(p_{0}+v, \kappa )] \nonumber \\ 
%& & -\frac{\left| p_{0} +v \right| + \sqrt{(p_{0}+ v)^2 - \kappa^2} + i {\rm sgn}[p_{0}+ v] \kappa
%}{2\sqrt{\left| p_{0} +v \right|\sqrt{(p_{0}+ v)^2 - \kappa^2}}}[C_{1}^{\searrow} G^{*}(p_{0}+v, \kappa ) + C_{2}^{\searrow} F(p_{0}+v, \kappa)] \nonumber
%\end{eqnarray}
%\end{widetext}
%\begin{widetext}
	\begin{eqnarray}
& & \frac{r^{\searrow}}{t^{\searrow}}e^{-i{\rm sgn}[p_{0}+v ] p_{0} \sqrt{(p_{0}+ v)^2 - \kappa^2}} = \nonumber \\	
	& & C_{1}^{\searrow}(v,\kappa )[\psi_{-}(p_{0}+v,\kappa )F^{*}(p_{0}+v, \kappa ) - \psi_{+}(p_{0}+v, \kappa)G^{*}(p_{0}+v, \kappa )] \nonumber \\
& &	+ C_{2}^{\searrow}[\psi_{-}(p_{0}+v,\kappa )G(p_{0}+v, \kappa ) - \psi_{+}(p_{0}+v, \kappa)F(p_{0}+v, \kappa)]. \label{rdown}
	\end{eqnarray}
%\end{widetext}
The absolute value of transmission is shown in Fig. \ref{fig:transmisssions} b).

\subsection{Symmetries}

\begin{figure}
	\includegraphics[width=0.99\columnwidth]{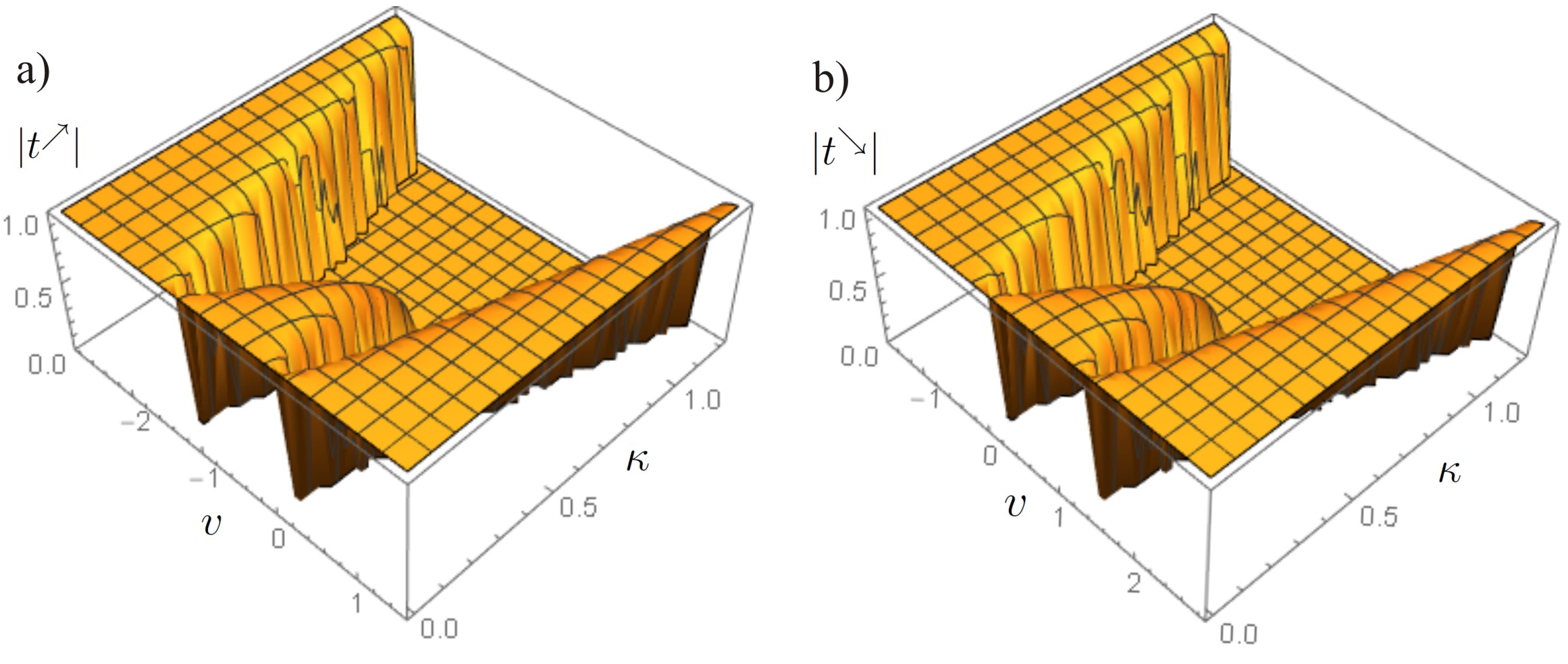}
	\caption{(color online) a) 	Transmission at the left side of the barrier for positive slope and impact parameter $p_{0}=1.3$. b) Transmission at the left side of the barrier for negative slope and impact parameter $p_{0}=-1.3$.}
	\label{fig:transmisssions}
\end{figure}

From the results shown in Fig. \ref{fig:transmisssions} we can already see a number on interesting features, which are detailed below. First, note that propagation is possible only for $\kappa$ satisfying $-v < \kappa < v$ and $-p_{0}-v <\kappa < p_{0} + v$, which implies zero transmission in regions that do not satisfy these conditions. Geometrically, the wave incident on the barrier in the metal region forms an angle $\phi_{\mathrm{M}}$ with the normal, and the transmitted wave in the graphene region forms an angle $\phi_{\mathrm{G}}$ with the normal. Here $\tan \phi_{\mathrm{M}} = \hbar k_{y}/p_{x, {\rm FM}} = 
\kappa /({\rm sgn}[p_{0}+v](\sqrt{(p_{0}+v)^2-\kappa^2})$ and  $\tan \phi_{\mathrm{G}} = \hbar k_{y}/p_{x, {\rm GM}}=\kappa /({\rm sgn}[v]\sqrt{v^2-\kappa^2})$. We have then an analogue of Snell's law, $v\sin  \phi_{\mathrm{G}} = (p_{0}+v)\sin\phi_{\mathrm{M}}$. We notice that the extreme values $p_{0}+v = \pm k$,
$v=\pm \kappa$ correspond to grazing incidence and transmission, that is $\phi_{\mathrm{M}} = \pi/2$
and $\phi_{\mathrm{G}} = \pi/2$ respectively.

\paragraph{Klein tunneling and pseudo-spin conservation}

We start by discussing the interesting case in which $\kappa=0$, {\it i.e.} the particle is incident on the barrier in the perpendicular $x$-direction. By analysing the final expression for transmission or reflection Eqs. (\ref{tup},\ref{rup}) and Eq. (\ref{tdown},\ref{rdown}), we can find immediately that the reflection is zero in this case, irrespective to the height or slope of the barrier. This is in contradistinction with the case of tunneling in non-relativistic quantum mechanics. It is instructive to understand what is the origin of this effect.

For $k=0$, the wavefunction under the metal reads
\begin{equation}
\Psi (\xi)_{x<d} = \frac{1}{t}\left[\begin{array}{c} 1 \\ 0 \end{array}\right] e^{i(p_{0} + v)\xi}+ \frac{r}{t}\left[\begin{array}{c} 0 \\ 1 \end{array}\right]e^{-i(p_{0} + v)\xi },
\end{equation}
the wavefunction on the top of the barrier  is
\begin{equation}
\Psi (\xi)_{x<d} = \left[\begin{array}{c} 1 \\ 0 \end{array}\right] e^{iv \xi},
\end{equation}
while in the connecting slope region we get $F(\xi - \xi_{0},0)= \exp\left(-i(\xi - \xi_{0})^{2}/2\right)$ and $G(\xi - \xi_{0},0)=0$ with solutions
\begin{equation}
\Psi^{\nearrow}(\xi )_{-d<x<0} = C_{1}^{\nearrow} \left[\begin{array}{c} 1\\ 0
\end{array} \right]e^{-i(\xi - \xi_{0})^{2}}
+
C_{2}^{\nearrow } \left[\begin{array}{c} 0 \\ 1
\end{array} \right] e^{i(\xi - \xi_{0})^{2}},
\end{equation}
for positive slope 
and 
\begin{equation}
\Psi^{\searrow}(\xi )_{-d<x<0} = C_{1}^{\searrow} \left[\begin{array}{c} 1\\ 0
\end{array} \right]e^{i(\xi - \xi_{0})^{2}} +
C_{2}^{\searrow } \left[\begin{array}{c} 0 \\ 1
\end{array} \right] e^{-i(\xi - \xi_{0})^{2}},
\end{equation}
for negative slope.
By imposing the continuity conditions we find immediately that $C_{1}^{\nearrow}=e^{iv^{2}/2}$, $C_{2}^{\nearrow }=0$, 
$C_{1}^{\searrow}=e^{-iv^{2}/2}$, $C_{2}^{\searrow }=0$, 
and $t^{\nearrow} = e^{-i p_{0}(p_{0}+v)} e^{+i (p_{0}+v)^{2}/2} e^{-i v^{2}/2}=e^{-ip_{0}^2/2}$, $r^{\nearrow}=0$,
 and $t^{\searrow} = e^{i p_{0}(p_{0}+v)} e^{-i (p_{0}+v)^{2}/2} e^{i v^{2}/2}=e^{ip_{0}^2/2}$, $r^{\searrow}=0$.  Thus, the transmission coefficient picks up a phase which is independent on energy and there is no backscattered wave. As mentioned before, this result can be obtained direcly from  Eqs. (\ref{tup},\ref{rup}) and Eq. (\ref{tdown},\ref{rdown}) but the explicit derivation above allows us to identify clearly the origin of this phenomenon. Indeed, this is now manifest as the conservation of pseudospin: the state at the top of the barrier is an eigenstate of the pseudospin $\sigma_{z}=+1$, which  forces all the components with $\psi_{-}(x)\neq 0$ on the slope and in the metal region to be zero.

\paragraph{Global symmetries}

The special symmetry of the trapezoidal potential with respect to half of its height leads to two additional symmetries of this problem. Note that while time-reversal and particle-hole are local symmetries, valid for a given $x$, the symmetries unearthed below are global symmetries, arising due to the shape of the solution across the slope and the connection points. In terms of energies, this amounts for the substitution  $p_{0}/2  + v\leftrightarrow -p_{0}/2 -v$ or 
$v \leftrightarrow -p_{0}-v$.

The first such symmetry is the following one:
\begin{eqnarray}
\frac{1}{t^{\nearrow}(-p_{0}-v, \kappa)}
e^{i {\rm sgn}[v]\sqrt{v^2 - \kappa^2}}&=&
\frac{1}{t^{\searrow}(v, \kappa)}e^{i {\rm sgn}[p_{0}+ v]\sqrt{(p_{0}+v)^2 - \kappa^2}},\\
\frac{r^{\nearrow}(-p_{0}-v, \kappa)}{t^{\nearrow}(-p_{0}-v, \kappa)}
e^{-i {\rm sgn}[v]\sqrt{v^2 - \kappa^2}}&=&
- \frac{r^{\searrow *}(v, \kappa)}{t^{\searrow *}(v, \kappa)}e^{i {\rm sgn}[p_{0}+ v]\sqrt{(p_{0}+v)^2 - \kappa^2}}.
\end{eqnarray}

Using the expressions for the left side of the barrier, one can immediately obtain the corresponding expressions for the right side. For example, a particle impinging on the barrier from the left side will experience on the right side a step with the opposite sign. Suppose that the slope at the left side is $p_{0}$: then to find the transmission coefficient at the right side it is enough to make the substitution $v\rightarrow p_{0} + v$ or $v-p_{0} \rightarrow v$ in the negative-slope formula (if the slope at  right side is negative) or in the positive-slope formula (if the slope at the right side is positive).

The second symmetry concerns the absolute value of the transmission coefficient, namely in both the case of positive slope and in that of negative slope the transmission coefficient is symmetric with respect to the point $-p_{0}/2$. This can be observed from Fig. \ref{fig:transmisssions}. So $|t(v,\kappa)|$ is invariant under the transformation $v \rightarrow -p_{0}-v$, in other words $|t(v,\kappa)|=|t(-p_{0}-v,\kappa)|$ and by consequence of normalization  $|r(v,\kappa)|=|r(-p_{0}-v,\kappa)|$.  We emphasize that this property is non-trivial: it not a direct consequence of one particular local symmetry of the problem but rather a consequence of the global inversion symmetry of the potential $V(x)$ with respect to the point where the energy $-p_{0}/2$ crosses the slope. We have also checked numerically that this symmetry does not extend to the argument of $t$, with the exception of the case $\kappa =0$ analysed in detail above. In the latter situation, this happens because the phase is anyway independent on $v$: it is $-p_{0}^2/2$ for the positive slope case and $p_{0}^2/2$ for the negative slope case. Using the code employed to plot Fig. \ref{fig:transmisssions}, we have also verified numerically that these are indeed the correct values of the phases.

\section{Conductance and Fano factor for the whole barrier}

With state-of-the-art nanofabrication of graphene junctions, it is possible to realize samples that are in the ballistic regime in the region of interest. Next, we work out in detail how the conductance and Fano factor can be calculated, allowing a straightforward comparison with experiments.

As established above, an incident wave coming from the left of the trapezoidal barrier will be partly reflected, with a complex reflection coefficient $r$, and partly transmitted, with transmission coefficient $t$. These coefficients will depend on $p_{0}$, $v$ and $\kappa$,
$r=r(v,\kappa)$, $t=t(v,\kappa)$, with each value of $k_{y}$ corresponding to a conduction channel. Therefore, the overall conductance and noise is a sum of the conductances and respectively noise per channel.

Specifically, the conductance is obtained as
\begin{equation}
\sigma (v) = \sigma_{0}\int_{0}^{{\rm min}\left\{\left|p_{0}+v\right|, \left|v\right|\right\}}
d\kappa T(v,\kappa),\label{gg}
\end{equation}
and the differential noise reads
\begin{equation}
s (v)= \sigma_{0}\int_{0}^{{\rm min}\left\{\left|p_{0}+v\right|,|v|\right\}}d\kappa [1-T(v,\kappa)]T(v,\kappa)\label{ss}.
\end{equation}
From this, we get the Fano factor as
\begin{equation}
F(v) = \frac{s(v)}{\sigma (v)},\label{FF}
\end{equation}
where $\sigma_{0}=4e^{2}w\sqrt{a}/\pi h = 2 G_{0}w\sqrt{a}/\pi$, with $w$ the width of the sample and $G_{0} = 2 e^{2}/h = 7.74\times 10^{-5}\Omega^{-1}$ the quantum of conductance. The factor of 2 that multiplies $G_{0}$ in the expression of $\sigma_{0}$ is due to the double-valley degeneracy of graphene. Note also the upper limits of the integrals in Eqs. (\ref{gg},\ref{FF}), 
which come from the condition of free propagation at the top of the barrier 
$\kappa <|v|$, and at the left side of the barrier $\kappa <\left|p_{0}+v\right|$.

An important observation concerns the role played by the barrier impact factor $p_{0} = \sqrt{a}d$ as
a fitting parameter.  For a given sample geometry and for a certain $V_g$, the Fano factor depends only on $p_0$. The gate voltage $V_g$ completely determines the value of $\Delta E_{\rm FG}$ and of $v = k_{\rm FG}/\sqrt{a}$. Then only $p_0$ remains in the expression of the Fano factor. Note that $p_0 = \sqrt{a}d$, and exactly this quantity appears in the arguments of the hypergeometric functions when we calculate $\xi (-d)$. This is not the case for $\sigma$ -- there the integral would depend only on $p_0$ but $\sigma_0$ contains another dependence on the slope a.

\subsection{Ballistic incoherent case}

\begin{figure}
\includegraphics[width=0.99\columnwidth]{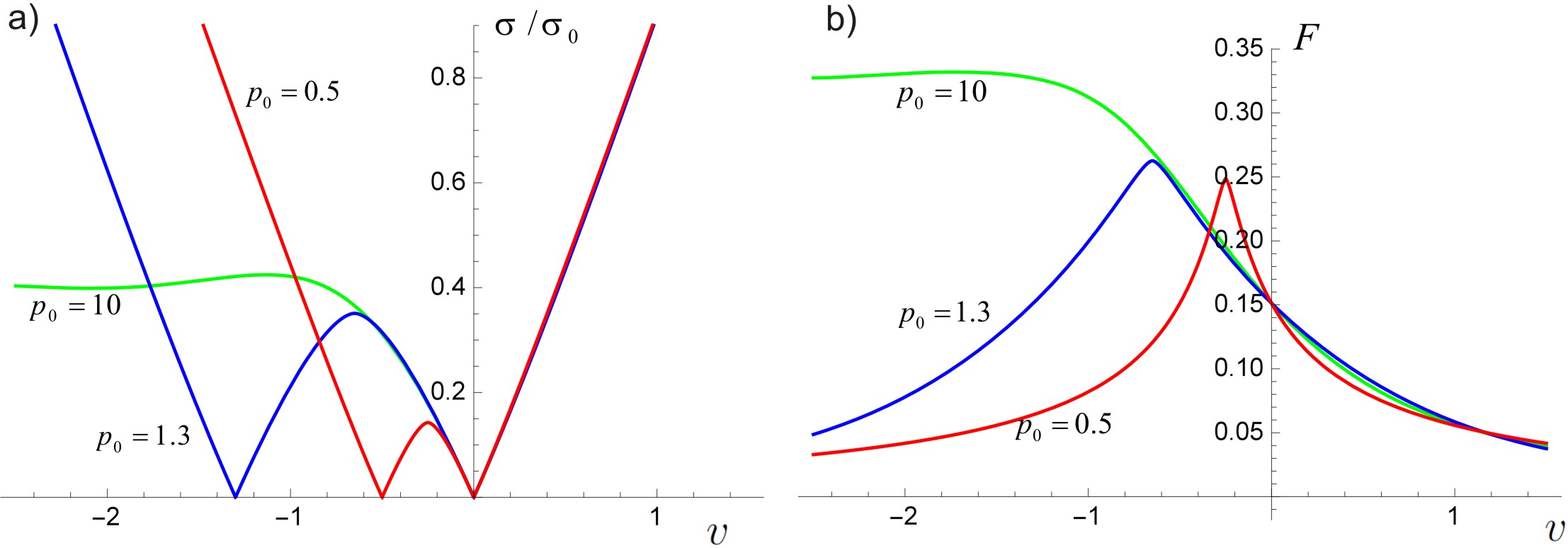}
\caption{(color online) Conductance a) and Fano factor b) for positive slope at the left of the barrier. The green lines correspond to $p_{0}=10$, the blue ones to $p_{0}=1.3$, and the red ones to $p_{0}=0.5$.}
\label{conductance_fano_leftpositivebarrier}
\end{figure}

\begin{figure}
\includegraphics[width=0.99\columnwidth]{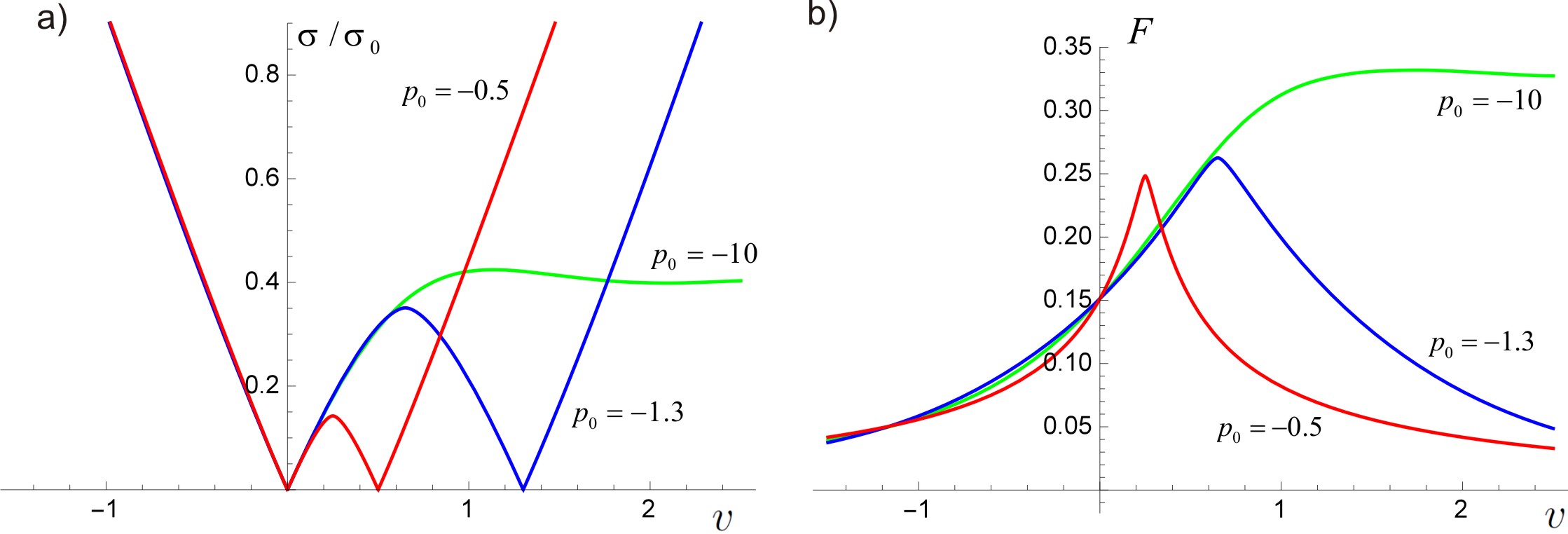}
\caption{(color online) Conductance a) and Fano factor b) for negative slope at the left of the barrier. The green lines correspond to $p_{0}=-10$, the blue ones to $p_{0}=-1.3$, and the red ones to $p_{0}=-0.5$.}
\label{conductance_fano_leftnegativebarrier}
\end{figure}

In the case of incoherent tunneling, the length of the sample is larger than the coherence length and as a result the phase factors appearing in the transmission is lost as the particle propagates on the top of the graphene barrier. 
This results in a transmission probability
\begin{equation}
T(v,\kappa )= \frac{1}{2|t(v,\kappa)|^{-2}-1}, \label{transmission}
\end{equation}
where we have considered a symmetric barrier (same slope at the left and right side) with reduced Fermi level $v$ in the region G. Using Eqs. (\ref{gg},\ref{FF}) we can calculate the conductance and the Fano factor for several values of the impact parameter, see Fig. \ref{conductance_fano_leftpositivebarrier} and Fig. \ref{conductance_fano_leftnegativebarrier}.

A number of interesting features can be noticed from the plots. For large impact parameter $|p_{0}|\gg 1$ we recover the previous results of Sonin \cite{sonin2009}. In practice, we see that it is sufficient to have $|p_{0}| \gtrsim 5$ to recover the main features corresponding to the limit  $|v|$ smaller that $|p_{0}|$. For $|p_{0}|=10$ we get the saturation conductance $\sigma /\sigma_{0} \approx 0.4$ and $F \approx 0.33$. 
As the parameter $p_{0}$ decreases, the conductance at low voltages tends to decrease, due to the fact that the transmission decreases. We no longer observe a saturation phenomenon, but instead, in the range $-|p_{0}|<|v|<-|p_{0}|$
the maximum values for the conductance and Fano factors are reached at $v=-p_{0}/2$. For values of $|p_{0}|$ below 1, the maximum value of the Fano factor is $\approx 0.25$, and it stays almost independent of $p_{0}$.
The slope of the conductance outside the range $-|p_{0}|<|v|<-|p_{0}|$ as well as the value of the Fano factor at $v=0$ and $v=p_0$
also take universal values, independent on $p_{0}$. We can find these values by looking at $0<v\ll 1$ for $p_{0}>0$ and at $0<-v\ll 1$ for $p_{0}<0$, and using $F(0,\kappa)=1$ and $G(0,\kappa)=1$, to get
\begin{equation}
|t (v,\kappa)|^2 \approx \frac{2 \sqrt{v^2-\kappa^2}}{|v|+\sqrt{v^2-\kappa^2}},
\end{equation}
and therefore
\begin{equation}
T \approx \frac{\sqrt{v^2-\kappa^2}}{|v|}.
\end{equation}
Inserting this expression into Eq. (\ref{gg}, \ref{FF}) we obtain by direct integration that
\begin{equation}
\frac{1}{\sigma_{0}}\frac{d\sigma}{d |v|} = \frac{\pi}{4} = 0.785,
\end{equation}
and
\begin{equation}
F(v=0) = 1 - \frac{8}{3 \pi} = 0.151 .
\end{equation}

\subsection{Ballistic coherent case}

\begin{figure}
	\includegraphics[width=0.99\columnwidth]{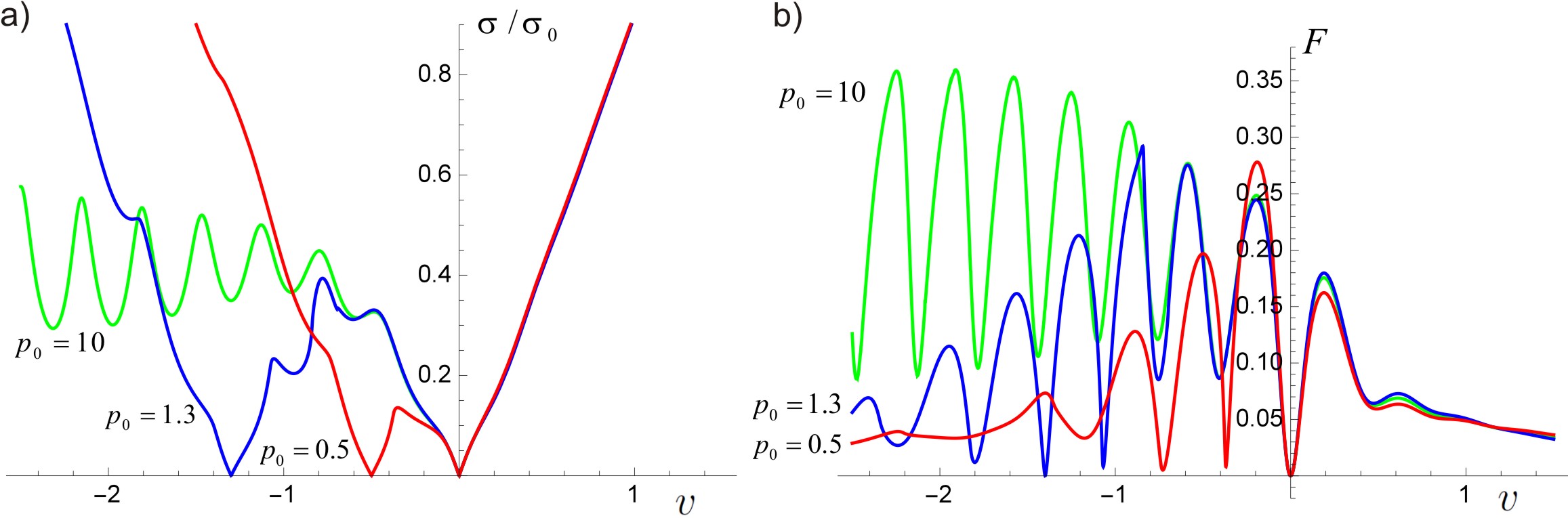}
	\caption{(color online) Conductance a) and Fano factor b) for positive slope at the left of the barrier in the coherent case. The green lines correspond to $p_{0}=10$, the blue ones to $p_{0}=1.3$, and the red ones to $p_{0}=0.5$.}
	\label{conductance_fano_leftpositivebarrier_coherent}
\end{figure}

If the propagation in the graphene sheet is coherent, then Fabry-P\'erot interference fringes will appear in the conductance and in the Fano factor due to interference between multiple reflections at the two slopes of the barrier. The formula for the total transmission in this case is
\begin{equation}
T = \frac{|t|^{4}}{1+|r|^{4}
	-2|r|^{2} \cos [ 2 \xi_{L}\sqrt{v^{2}-\kappa^2}]}, \label{cohe}
\end{equation}
where we used the fact that the wave propagating from the left slope to the right slope in the non-contact region picks up a phase $\sqrt{v^{2}-\kappa^2}\xi_{L}$, where  $\xi_{L}$ is the reduced length of the top of the barrier, $\xi_{L} = \sqrt{a}L$. This result can be employed now in Eq. (\ref{gg}, \ref{FF}) to calculate the conductance and the Fano factor. As an example, in the case of positive slope at the left side of the barier the conductance and the Fano factor are plotted in Fig. \ref{conductance_fano_leftpositivebarrier_coherent} with $\xi_{L} =\sqrt{a}L = 9$. For ballistic and coherent graphene samples, the modulation of conductance and Fano factor around the Dirac point $v=0$ has been already observed experimentally in Fabry-P\'erot experiments  \cite{Pertti}. One notices that in the conductance the modulation due to multiple reflections is not so pronounced for $v>0$. This is another manifestation of the Klein tunneling phenomenon, and it is due to the fact that in that region the transmission is large, therefore there exists comparatively little reflection with respect to the negative-$v$ region. In Eq. (\ref{cohe}) the effect of the modulation with respect to $\xi_{L}$ thus tends to be comparatively small and it is further washed out by the integration over $\kappa$. Thus the Fabry-P\'erot effect tends to be suppressed, and only the Fano factor retains some mild oscillatory dependence.

%%%%%%%%%%

\section{Experimental proposal}

To test these prediction, a rather standard experimental setup, together with high-quality clean graphene and metal-graphene contacts is needed. These are available nowadays, owing to  progress in the chemical vapor deposition synthesis of graphene on metal surfaces \cite{Wintterlin} and novel techniques for achieving a controlled small contact resistance \cite{Cusati}. The ballistic regime, with electrons having long mean-free paths compared to the sample size, can be achieved by fabrication methods for suspended graphene using either conventional wet etching of the SiO$_2$ substrate \cite{Andrei} or by dry transfer \cite{Pertti}. Top gating is also possible, for example by using a solid polymer electrolyte \cite{ferrari}. These methods ensure that the electrons have mean free paths larger than the dimensions of the sample \cite{kim}. Several calculations have confirmed that the details of the contact do not alter significantly the ballistic regime and only slight modifications of the values of the conductance and Fano factor are to be expected  
 	\cite{observations1,observations2}. We have also verified (see Appendix \ref{insensitive}) that our experimental predictions are largely insensitive to the details of tunneling between metal and graphene. This makes our results robust against uncontrolled fabrication imperfections.

We introduce next a doping model that allows us to determine the renormalized Fermi level $v$ and the impact parameter $p_0$ from the geometric characteristics of the sample and the applied gate and bias voltages. The notations used are as follows: the workfunction of the pristine graphene is denoted by $W_{\rm G}$, the workfunction for the metal-covered region is $W_{\rm M}$, the workfunction of the backgate is $W_{g}$. The chemical shift \cite{giovannetti} of the workfunction of graphene at the contact with the metal this workfunction is denoted by $\Delta_c$.

\begin{figure}
	\includegraphics[width=0.97\columnwidth]{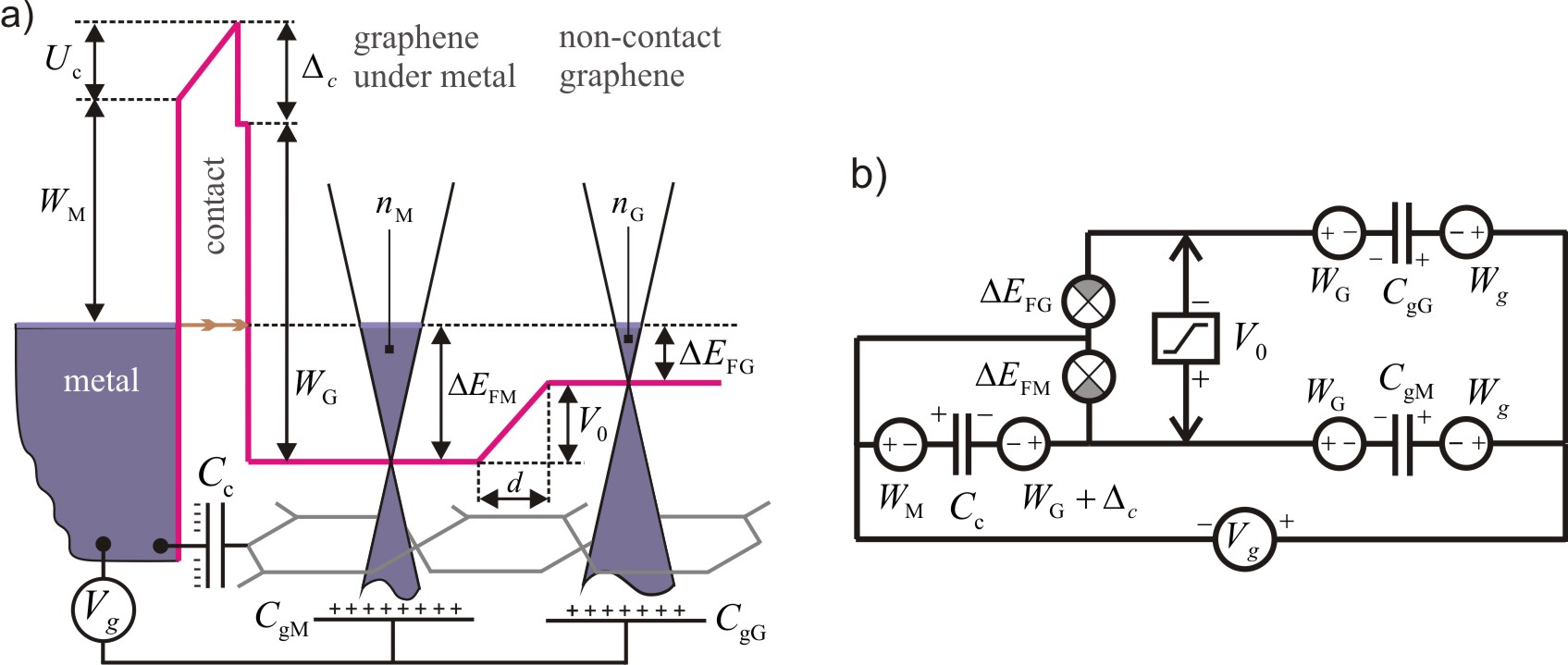}
	\caption{a) Schematic of Fermi levels for the doping problem. b) Equivalent electrical circuit corresponding to the electrochemical potentials.}
	\label{doping_schematic}
\end{figure}

With these notations, we can now find the shifts of the Fermi level in the region under the metal $\Delta E_{\rm FM}$ and in the non-contact region $\Delta E_{\rm FG}$ as a function of the applied gate voltage $V_{g}$.
In Fig.~\ref{doping_schematic} b) we present a full equivalent electric schematic of the circuit. From this, all the equations above (and combinations thereof) can be obtained by applying Kichhoff's laws for electrical circuits. The quantum capacitances corresponding to the graphene under the metal and the suspended graphene are represented by circles containing the Dirac cone. The differential conductance and the noise can be measured using a lock-in technique, with a small excitation voltage of a couple of tens of mV \cite{us}. This is much smaller compared to the other voltages and energy scales in this circuit, and therefore will be neglected in the following.

The circuit equations can be obtained from Kirchhoff's laws along three independent contours,
\begin{eqnarray}
&&e  V_{g} = W_{g} + \Delta_{c} - W_{\rm M} + e U_{g{\rm M}} - e U_{c}, \label{fm1} \\
&& 0 = \Delta E_{\rm FM} + e U_{c} - W_{\rm G} - \Delta_{c} + W_{\rm M}, \label{fm2} \\
&& e V_{g}=e U_{g{\rm G}} + \Delta E_{\rm FG}+ W_{g}- W_{\rm G} .  \label{fm3}
\end{eqnarray}
Here $U_{c}/e = n_{c}/C_{c}$ is the voltage across the surface capacitance $C_{c}$, with $n_{c}$ the number of electrons per unit area in the contact region, and 
$U_{g{\rm M}}/e = n_{g{\rm M}}/C_{g{\rm M}}$
is the voltage across $C_{g{\rm M}}$, corresponding to an electron distribution $n_{g{\rm M}}$ on the backgate. 
In the non-contact region, the surface distribution of electrons $n_{g{\rm G}}$ produces a voltage drop $U_{g{\rm G}}=en_{g{\rm G}}/C_{g{\rm G}}$.  Denoting the surface electron density in the graphene layer in the metal region by $n_{\rm M}$ and the surface density on the graphene side by $n_{\rm G}$, the usual laws of charge conservation yield
\begin{eqnarray}
n_{\rm M} &=& n_{g{\rm M}} + n_{c}, \label{f1} \\
n_{\rm G} &=& n_{g{\rm G}}. \label{f2}
\end{eqnarray}

The existence of excess charge carriers in the regions M and G can be related to the values of the corresponding Fermi levels by the relations
\begin{eqnarray}
\Delta E_{\rm FM} &=& \hbar v_{\rm F}{\rm sgn \left[n_{\rm  M}\right]}\sqrt{\pi |n_{\rm  M}|},\label{fg1} \\ 
\Delta E_{\rm FG} &=& \hbar v_{\rm F}{\rm sgn \left[n_{\rm  G}\right]}\sqrt{\pi |n_{\rm  G}|}.  \label{fg2}
\end{eqnarray}

From Eqs. (\ref{fm1}-\ref{fm3}), Eqs. (\ref{f1}-\ref{f2}) and Eqs. (\ref{fg1}-\ref{fg2}) we
can get now expressions for the Fermi level shifts 
\begin{widetext}
\begin{equation}
\Delta E_{\rm FM} = {\rm sgn}\left[\delta V_{g} + \frac{\chi C_{c}}{eC_{g{\rm M}}}\right]
\left\{-\frac{C_{c}+C_{g{\rm M}}}{2}\zeta_{\rm F}^{2}+\sqrt{\left(\frac{C_{c}+C_{g{\rm M}}}{2}\zeta_{\rm F}^{2}\right)^2
+\zeta_{\rm F}^2 C_{c} \left\vert \chi + \frac{C_{g{\rm M}}}{C_c}
e\delta V_{g}
\right\vert}\right\}, \label{sol1}
\end{equation}
\begin{equation}
\Delta E_{\rm FG} = {\rm sgn}\left[\delta V_{g}\right]
\left\{-\frac{C_{\rm gG}}{2}\zeta_{\rm F}^{2}+\sqrt{\left(\frac{C_{\rm gG}}{2}\zeta_{\rm F}^{2}\right)^2
+ e C_{\rm gG}\zeta_{\rm F}^2\vert \delta V_{g}\vert}\right\}.\label{sol2}
\end{equation}
\end{widetext}
which are directly connected to the experimentally-relevant quantities. Here $\zeta_{\rm F}= \sqrt{\pi}\hbar v_{\rm F}/e$ is a constant called Fermi electric flux \cite{us}, and the value $v_{\rm F} = 1.1 \times 10^8$ cm/s for the Fermi velocity of graphene yields $\hbar v_{\rm F} = 0.724$ eV$\cdot$ nm and $\zeta_{\rm F} = 1.283 \times 10^{-7}$ V$\cdot$cm.
We also defined $\delta V_{g} = V_{g} - (W_{g}- W_{\rm G})/e$, and
the difference $\chi = W_{\rm G} + \Delta_{c} - W_{\rm M}$ between the workfunction of the graphene in the contact region and the workfunction of the metal.

Typically the experiments as well the theory \cite{sonin2009} have been focused on observing the Dirac point corresponding to the suspended graphene region, achieved for $\Delta E_{\rm FG}=0$, see Eq. (\ref{sol2}). We show here that given a right combination of sample parameters, also the additional Dirac point of graphene under metal corresponding to $\Delta E_{\rm FM}=0$ is accessible. This requires 
\begin{equation}
C_{c} \chi + C_{g{\rm M}}e \delta V_{g} = 0, \label{eq:seconddirac}
\end{equation}
see Eq. (\ref{sol1}). As a rule of thumb, this means that the contribution of the doping and contact coupling in this equation should not be too large, such that $\delta V_{g}$ is still within the measurable range.

\begin{figure}
	\includegraphics[width=0.99\columnwidth]{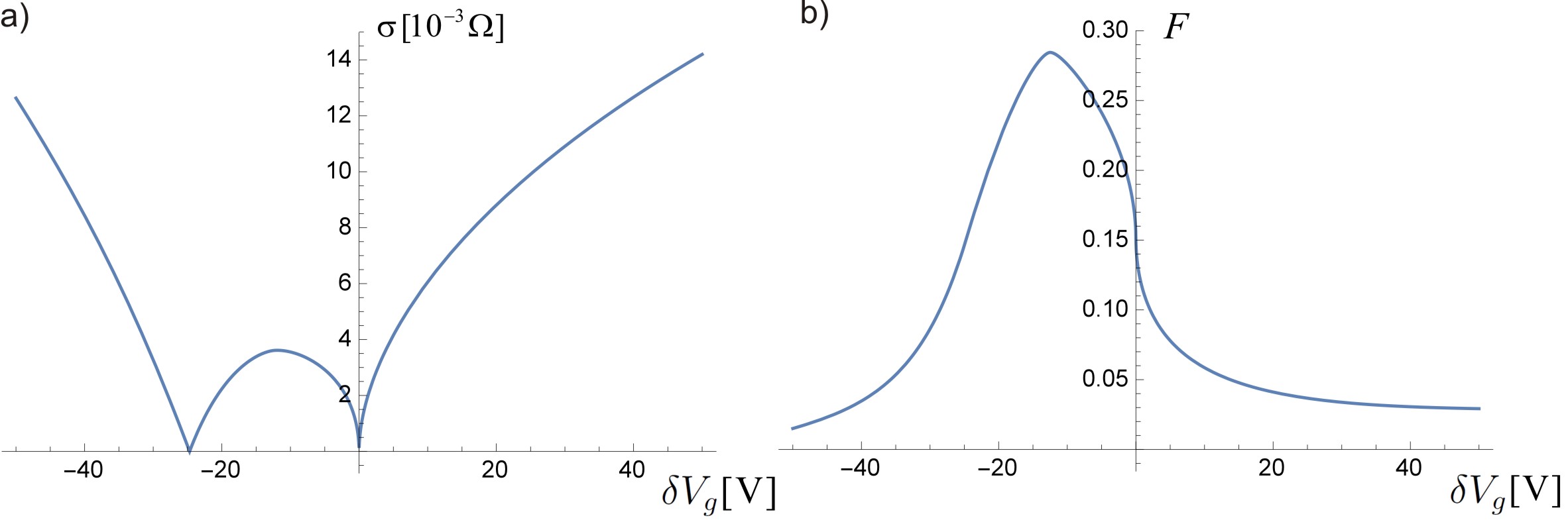}
	\caption{a) Conductance and b) Fano factor for the sample discussed in the text. For this numerical simulation we took $d=10$ nm, $\chi = 0.3$ eV  and a sample width $w=1$ $\mu$m. For the capacitances per unit area we took $C_{c}= 2.9 \times 10^{-6}$ F/cm$^{2}$, $C_{g{\rm G}}= 8.8 \times 10^{-9}$ F/cm$^{2}$, and $C_{g{\rm M}}= 3.4 \times 10^{-8}$ F/cm$^{2}$.}
	\label{with_dopingmodel}
\end{figure}

We can then connect the experimental parameters with the barrier parameter $p_0$ and the effective Fermi level $v$ by
\begin{equation}
eV_{0} = \Delta E_{\rm FM}-\Delta E_{\rm FG}.
\end{equation}
Using $p_{0}=eV_{0}/\hbar v_{\rm F}\sqrt{a}$ we obtain
\begin{widetext}
\begin{eqnarray}
v&=&\left( \frac{d}{\hbar v_{\rm F}}\right)^{1/2}\frac{\Delta E_{\rm FG}}{\sqrt{\vert\Delta E_{\rm FM}-\Delta E_{\rm FG}\vert}} =
\left( \frac{\sqrt{\pi}d}{e \xi_{\rm F}}\right)^{1/2}\frac{\Delta E_{\rm FG}}{\sqrt{\vert\Delta E_{\rm FM}-\Delta E_{\rm FG}\vert}}, \label{vee} \\
p_{0}&=& {\rm sgn}[V_{0}]\left( \frac{d}{\hbar v_{\rm F}}\right)^{1/2}
\sqrt{\vert\Delta E_{\rm FM}-\Delta E_{\rm FG}\vert} = {\rm sgn}[V_{0}]\left( \frac{\sqrt{\pi}d}{e \xi_{\rm F}}\right)^{1/2}\sqrt{\vert\Delta E_{\rm FM}-\Delta E_{\rm FG}\vert}.\label{pee}
\end{eqnarray}
\end{widetext}
The slope of the barrier can be expressed as well in terms of the voltage-dependent $\Delta E_{\rm FM}$ and  $\Delta E_{\rm FG}$,
\begin{equation}
a= \frac{1}{\hbar v_{\rm F}d}\vert\Delta E_{\rm FM}-\Delta E_{\rm FG}\vert =
\frac{\sqrt{\pi}}{e\xi_{\rm F}d}\vert\Delta E_{\rm FM}-\Delta E_{\rm FG}\vert,\label{aaa}
\end{equation}
and we also have $p_{0}={\rm sgn}[V_{0}]d\sqrt{a}$. We have naturally assumed here that the distance $d$ over which the electric potential drop $V_{0}$ occurs is constant (does not change with the gate voltage). However, we have verified numerically that if one takes the slope $a$ as being constant, there are only minor quantitative changes in the results. This shows that our results are robust as they are insensitive to the details of the chemistry of the rearrangement of carriers at the electrostatic barrier.

For the experimental observation of the effects predicted before, the sample can be designed following relatively standard methods. The choice of the contact metal is very important. It is known that some metals (Co, Ni, Ti, Pd) interact very strongly with graphene, resulting in hybridization between the graphene $p_{z}$ orbitals and the electronic $d$ states in the metal. This chemisorption process destroys the Dirac cone. However, for metals such as Au, Ag, Cu, Al and Pt(111) the interaction is weaker (physisorption) and the cone is preserved \cite{giovannetti,Khomyakov,Zheng}. For the latter metals, it was also found by ab-initio calculations that the dangling bonds from either the unsaturated graphene edge or point defects further lowers the contact resistance, especially for Au and Ag \cite{Masuda,Bo}. From various experiments and theoretical considerations it is reasonable to expect a value of $d$ of around 10 nm, and this is the value we use in the simulations below.  The contact capacitance $C_{c}$ can be estimated as 
$C_{c}= 2.9 \times 10^{-6}$ F/cm$^{2}$ by assuming a equilibrium distance value of about 3 \AA ~between the graphene and the metal, 
%$C_{c}= 2.95 \times 10^{-6}$ F/cm$^{2}$ by assuming a equilibrium distance value of about 3 \AA ~between the graphene and the metal, 
%$C_{c}= 2.68 \times 10^{-6}$ F/cm$^{2}$ by assuming a equilibrium distance value of about 3.3 \AA ~between the graphene and the metal, 
as found from DFT calculations \cite{giovannetti}
for metals such as Al, Ag, Cu, Au, and Pt. From the same calculations, the workfunction of uncontacted graphene results as $W_{\rm G}=4.5$ eV, and the chemical interaction between graphene and metal for the equilibrium distance produces a shift $\Delta_{c} \approx 0.9$ eV \cite{giovannetti}. Since the workfunctions $W_{\rm M}$ of metals are typically also of the order of 1 eV closer to the pristine graphene, it is reasonable to consider for exemplification a metal-graphene workfunction difference $\chi = W_{\rm G} + \Delta_{c} - W_{\rm M}   = 0.3$ eV, very close to the value expected from a graphene-on-copper sample \cite{giovannetti, Khomyakov}. Then, let us take a sample with 100 nm separation of the graphene layer from the backgate, with an air gap in the region of non-contact and with SiO$_2$ supports (relative permittivity $\epsilon_{r}=3.9$) in the contact region. 
%This results in $C_{g{\rm G}}= 9.2 \times 10^{-9}$ F/cm$^{2}$ and $C_{g{\rm M}}= 1.2 \times 10^{-8}$ F/cm$^{2}$.???
This results in $C_{g{\rm G}}= 8.8 \times 10^{-9}$ F/cm$^{2}$ and $C_{g{\rm M}}= 3.4 \times 10^{-8}$ F/cm$^{2}$.

The numerical calculations for the conductance and the Fano factors are plotted in Fig. \ref{with_dopingmodel}.
The conductance and the Fano factor appear slightly deformed when compared with the representation in terms of the reduced Fermi momentum $v$ presented in the previous sections, due to the fact that Eqs. (\ref{sol1},\ref{sol2}) and
Eqs. (\ref{vee},\ref{pee}) are nonlinear, but the key features are clearly there. The position of the additional Dirac point is obtained numerically at $\delta V_{g} =-25.6$V, in agreement with Eq. (\ref{eq:seconddirac}).

We note also that most graphene samples present corrugations: these effects are equivalent to an electric field perpendicular to the surface of the sample or to a magnetic field along the sheet. However, due to the fact that the spin-orbit interaction is small in graphene, the resultant Rashba splitting for curvature radii of around 100 nm is about 17 $\mu$eV (0.2 K) \cite{Brataas}, much smaller than the other energies ({\it e.g.} the hopping constant in graphene is 2.8 eV and the workfunctions are of the order of a few eV), and therefore can be neglected. Also the presence of evanescent modes can be neglected for bias voltages larger than $\zeta_{\rm F}/L = \sqrt{\pi}\hbar v_{\rm F}/(eL)$, where $L$ is the length of the sample \cite{sonin2008}. For $L=1$ $\mu$m, this corresponds to voltages of the order of $1.283$ mV, which is easily satisfied experimentally.

\section{Discussions and conclusions}

In elementary quantum mechanics, tunneling under the trapezoidal barrier is a paradigmatic problem.
Yet to date, in the case of graphene this problems has not been completely solved, with only partial solutions being presented in the literature. Here we present the full solution of the problem of Klein tunneling though the trapezoidal barrier. We calculate the transmission and reflection coefficients
and describe their symmetries. By using these coefficients, we can further proceed and calculate the conductance, the noise, and the Fano factor. For realistic experimental setups with contacts realized from metals with certain workfunctions, we present a model for doping that allows us to make specific predictions that can be verified experimentally.

\section{Acknowledgements}
I acknowledge inspiring discussions with Edouard Sonin on the symmetries of the problem as well as with Pertti Hakonen, Bernard Pla\c cais, and Antti Laitinen on possible experimental verifications. 
This work was supported  by the “Finnish Center of Excellence in
Quantum Technology QTF” (project 312296) of the Academy of Finland. We also acknowledge the European Microkelvin Platform (EMP, No. 824109).

\appendix

\section{Derivation of the Sauter solutions}
\label{kummer}

For completeness, we provide here an explicit derivation of the solutions for the Dirac equation in the positive slope region. From Eq. (\ref{eq:Dirac}) with the kernel ${\cal K}(x) = -{\rm sgn} [V_{0}] a (x - x_{0})$ we get for each of the pseudospin components
\begin{eqnarray}
i\frac{d}{dx}\psi_{+}(x) + i k_{y}\psi_{-}(x) &=& {\rm sgn} [V_{0}] a (x - x_{0})\psi_{+}(x), \\
i\frac{d}{dx}\psi_{-}(x) + i k_{y}\psi_{+}(x) &=& -{\rm sgn} [V_{0}] a (x - x_{0})\psi_{-}(x).
\end{eqnarray}
By employing the notations $\kappa =k_{y}/\sqrt{a}$, $\xi =\sqrt{a}x$, and $\xi_{0} =\sqrt{a}x_{0}$
we can further write 
\begin{eqnarray}
\left[\frac{d}{d\xi} + i {\rm sgn} [V_{0}] (\xi-\xi_{0})\right]\psi_{+}(\xi ) + \kappa \psi_{-}(\xi ) &=& 0, \label{euph}\\
\left[\frac{d}{d\xi} - i {\rm sgn} [V_{0}] (\xi-\xi_{0})\right]\psi_{-}(\xi ) + \kappa \psi_{+}(\xi )&=& 0, \label{neeuph}
\end{eqnarray}
see also Eq. (\ref{11}) and Eq. (\ref{12}).
Combining these two equations we obtain
\begin{equation}
\left[\frac{d^{2}}{d\xi^{2}} + (\xi-\xi_{0})^2\right]\psi_{\pm}(\xi ) + \left(\pm i {\rm sgn}[V_{0}] - \kappa^2\right) \psi_{\pm}(\xi ) = 0.
\label{equationnn}
\end{equation}
Let us consider the case of positive slope, ${\rm sgn}[V_{0}] =+1$. First, we will find the solution $\psi_{+}^{\nearrow}(\xi )$. We proceed by making the substitution $z=i(\xi-\xi_{0})^2$ and $\psi_{+}^{\nearrow} (\xi )= \exp [-z/2 ] h (z )$ into Eq. (\ref{equationnn}), obtaining
\begin{equation}
z \frac{d^2}{dz^2}h(z) + \left(\frac{1}{2}-z\right) \frac{dh}{dz}+ \frac{i\kappa^2}{4} h (z) = 0.
\end{equation}
We can recognize this as Kummer's equation, with one solution denoted by $M\left(-i\kappa^2 /4, 1/2,z\right)$. This produces a solution for $\psi_{+}^{\nearrow}(\xi )$ as the function F defined in the main text,
\begin{equation}
F (\xi - \xi_{0}) = e^{-i(\xi - \xi_{0})^2/2}M\left[-\frac{i\kappa^2}{4}, \frac{1}{2}, i (\xi - \xi_{0})^2\right]. \label{eq:psiplus}
\end{equation}
In order to find the corresponding spinor component $\psi_{-}^{\nearrow} (\xi )$, which we will denote by $G$,  we can use directly Eq. (\ref{euph}),
\begin{equation}
G (\xi -\xi_{0}) = -\frac{1}{\kappa}\left[\frac{d}{d\xi} + i(\xi - \xi_{0})\right] F (\xi - \xi_{0}).
\end{equation}
Using the expression for the derivative of the Kummer function Eq. (13.4.8) from Abramowitz and Stegun \cite{Abramo}
\begin{equation}
\frac{d}{dz}M(a,b,z) = \frac{a}{b}M(a+1,b+1,z)
\end{equation}
we get
\begin{equation}
 \psi_{-}^{\nearrow}(\xi ) =  G(\xi - \xi_{0}) = -\kappa (\xi - \xi_{0}) e^{-i(\xi - \xi_{0})^2/2}M\left[1-\frac{i\kappa^2}{4}, \frac{3}{2}, i (\xi - \xi_{0})^2\right]. \label{eq:psiminus}
\end{equation}
As a consistency check, we can now verify the other equation  Eq. (\ref{neeuph}) with the help of the recurrence relation 13.4.14 from Abramowitz and Stegun \cite{Abramo},
\begin{equation}
(b-1) M(a-1, b-1, z) = (b-1-z)M(a,b,z) + z \frac{d}{dz} M(a,b,z) .
\end{equation}

We have now identified one solution for the case of positive slope, with pseudospin components  $\psi_{\pm}^{\nearrow}(\xi )$ given by Eqs. (\ref{eq:psiplus},\ref{eq:psiminus}). The other solution for positive slope can be obtained by using the electron-hole symmetry. If we take the complex conjugate of Eq. (\ref{euph},\ref{neeuph}) we find the same system of equations but with
$(\psi_{-}^{\nearrow})^{*}$ and $(\psi_{+}^{\nearrow})^{*}$ playing the roles of $\psi_{+}^{\nearrow}$ and $\psi_{-}^{\nearrow}$ respectively. Thus, we have another independent solution, $G^{*} (\xi -\xi_{0})$ for the $+$ component and $F^{*}(\xi - \xi_{0})$ for the $-$ component.

To conclude, for positive slope we have the general form of the solutions
\begin{equation}
\left[\begin{array}{c} \psi_{+}^{\nearrow}(\xi ) \\ \psi_{-}^{\nearrow}(\xi ) \end{array} \right]
=C_{1}^{\nearrow} \left[\begin{array}{c} F(\xi -\xi_{0}, \kappa) \\ G(\xi -\xi_{0},\kappa )
 \end{array} \right]
 +
 C_{2}^{\nearrow} \left[\begin{array}{c} G^{*}(\xi - \xi_{0}, \kappa) \\ F^{*}(\xi - \xi_{0}, \kappa )
 \end{array} \right].
 \end{equation}
Now, for the case of negative slope we can find the solutions by noticing that in Eqs. (\ref{euph},\ref{neeuph}) the negative-slope solutions can be obtained from the positive-sign solution by complex conjugation. We then have
\begin{equation}
\left[\begin{array}{c} \psi_{+}^{\searrow}(\xi ) \\ \psi_{-}^{\searrow}(\xi ) \end{array} \right]
=C_{1}^{\searrow} \left[\begin{array}{c} F^{*}(\xi -\xi_{0}, \kappa) \\ G^{*}(\xi -\xi_{0},\kappa )
 \end{array} \right]
 +
 C_{2}^{\searrow} \left[\begin{array}{c} G(\xi - \xi_{0}, \kappa) \\ F(\xi - \xi_{0}, \kappa )
 \end{array} \right].
 \end{equation}

\section{Insensitivity to details of the contact}
\label{insensitive}

The question to be  address in the following concerns the interface between the metal and the graphene sheet. We would like to investigate how the tunneling probability depends on the microscopic details of the tunneling between the metal and the graphene sheet.

To do so, let us consider that the electrons arriving from the metal side at the metal-graphene contact have only a finite transmission probability $T_{\rm c}$ of crossing this interface into graphene. If this process is incoherent, we have, based on general considerations (see e.g. Appendix B of Ref. \cite{us} for a pedagogical derivation) that the total transmission probability Eq. (\ref{transmission}) over the whole structure, becomes
\begin{equation}
T_{\rm tot}(v,\kappa )= \frac{1}{2|t_{\rm tot}(v,\kappa)|^{-2}-1}, \label{eq:1}
\end{equation}
where $t_{\rm L}(v,\kappa)$ is the one-side transmission probability as seen from the metallic leads,
\begin{equation}
\frac{1}{|t_{\rm tot}(v,\kappa)|^2} = \frac{1}{|t(v,\kappa)|^2} + \frac{1}{T_{\rm c}(v,\kappa)} -1 . \label{eq:2}
\end{equation}
We have now encapsulated the details of the interface into an unknown channel transmission probability $T_{\rm c}(v,\kappa)$.

In the absence of a detailed microscopic model, by running the simulations with various constant values of $T_{\rm c}(v,\kappa )=T_{\rm c}$, we can verify that the symmetry properties of the transmission probability are unaltered; the only effect of $T_{\rm c}$ is to reduce the maximum value of $\sigma$ and increase the maximum value of $F$ in Fig. 9, but otherwise the shape of these traces are very similar. 

In principle, one can now make one more step forward in this analysis and start to include assumptions about the interface, which may result in an energy-dependent $T_{c}$. For example, we can consider a simple model for the metal as a graphene sheet with a very low Dirac point, %\cite{sonin2008,sonin2009}
and assume that the electrons are injected into the graphene at some contact point $x_{c} < -d$. The wavefunction in the metallic leads then reads  
\begin{equation}
%\Psi (\xi)_{x <x_{\rm c}} = 
\frac{1}{t_{\rm c}} \left[\begin{array}{c} 1\\ 0 \end{array}\right] e^{iv_{\rm c}\xi} +
\frac{r_{\rm c}}{t_{\rm c}}\left[\begin{array}{c} 0\\ -1\end{array}\right] e^{- iv_{\rm c}\xi},\label{sp}
\end{equation}
where $v_{\rm c}$ is the reduced Fermi vector of the metal, assuming a relatively large value $v_{\rm c} \gg \kappa$. By matching Eq. (\ref{sp}) at $x_{\rm c} = \xi_{\rm c}/\sqrt{a}$  with  the transmitted wave 
\begin{equation}
\left[\begin{array}{c} \psi_{+}(p_{0}+v,\kappa)\\ \psi_{-}(p_{0}+v,\kappa) \end{array}\right] e^{i {\rm sgn}[p_{0}+v]\sqrt{(p_{0}+v)^2 - \kappa^2}\xi}
\end{equation}
at the right side of the contact (corresponding to the graphene sheet on the metal) 
we obtain 
\begin{equation}
\frac{1}{|t_{\rm c}(v,\kappa )|^2} = \frac{1}{T_{\rm c}(v,\kappa )}= |\psi_{+}(p_{0}+v,\kappa)|^2 = \frac{|p_{0}+v|}{2\sqrt{(p_{0}+v)^2-\kappa^{2}}}+\frac{1}{2}.  \label{countach}
\end{equation}
One can immediately see that $T_{\rm c}\leq 1$, and also that $T_{\rm c}\approx 1$ when $\kappa \ll |p_{0}+v|$. Note also that this contact model does not introduce any additional phenomenological parameters; in particular, owing to the assumption of incoherent tunneling, it does not depend on $\xi_{\rm c}$. By using Eq. (\ref{countach}) and Eqs. (\ref{eq:1},\ref{eq:2}) we can calculate the conductivity and Fano factor. We have verified numerically that, even under this energy-dependent model, the traces for conductance and shot noise do not change significantly for samples with parameters as used in this work.

Thus, we conclude that the main features of our model of tunneling under the trapezoidal barrier remain remarkably robust when the microscopic details of the interface between the metal and graphene are included.

%%%%%%%%%%%%%%%%%%%%%%%%%%%%%%%%%%%%%%%%%%%%%%%%%%%%%%%%%%%%%%

\end{document}